\documentstyle[11pt,amsfonts,fullpage]{amsart}

\newtheorem{theorem}{Theorem}[section]

\newtheorem{corollary}[theorem]{Corollary}

\newtheorem{definition}[theorem]{Definition}

\newtheorem{lemma}[theorem]{Lemma}

\newtheorem{proposition}[theorem]{Proposition}

\newtheorem{remark}[theorem]{Remark}

\def\endproof{\qed\smallskip}
\def\blacksquare{\hbox to .60em{\vrule width .60em height .60em}}

\newcommand{\bbgin}{\begin}

\begin{document}

\title[]{On Long-Time Evolution in General Relativity and Geometrization of 3-Manifolds}

\author[]{Michael T. Anderson}

\thanks{Partially supported by NSF Grant DMS 9802722}

\maketitle

\tableofcontents

\setcounter{section}{-1}

\section{Introduction.}

\setcounter{equation}{0}

 In this paper, we describe certain relations between the vacuum Einstein evolution equations in general relativity and the geometrization of 3-manifolds. In its simplest terms, these relations arise by analysing the long-time asymptotic behavior of natural space-like hypersurfaces $\Sigma_{\tau},$ diffeomorphic to a fixed $\Sigma ,$ in a vacuum space-time. In the best circumstances, the induced asymptotic geometry on $\Sigma_{\tau}$ would implement the geometrization of the 3-manifold $\Sigma .$

 We present several results illustrating this relationship, some of which require however rather strong hypotheses. Thus, besides proving these results, a second purpose of this paper is to clarify and make precise what some of the major difficulties are in carrying this program out to a deeper level. In this setting, we thus discuss a number of open problems and conjectures, some of which are well-known, which are of interest both in general relativity and in 3-manifold geometry. 

 The idea of geometrizing 3-manifolds, and its proof in many important cases, is due to Thurston. We refer to [38] for an introduction from the point of view of hyperbolic geometry. A survey of geometrization from the point of view of general Riemannian geometries is given in [1], where some early relations with issues in general relativity were also presented. We also refer to [6] and [34] for surveys on topics in general relativity related to this paper.

 Let ({\bf M, g}) be a vacuum space-time, i.e. a 4-manifold {\bf M}
 with smooth $(C^{\infty})$ Lorentz metric {\bf g}, of signature $(- , +,+,+),$ satisfying the Einstein vacuum equations
\begin{equation} \label{e0.1}
{\bf Ric_{g}}
 = 0. 
\end{equation}

 We assume that ({\bf M, g}) is a cosmological space-time in the sense that {\bf M} admits a compact Cauchy surface $\Sigma .$ In particular, ({\bf M, g}) is then globally hyperbolic and {\bf M}
  is diffeomorphic to $\Sigma \times {\Bbb R} .$ The space-time ({\bf M, g}) is the maximal $C^{\infty}$ vacuum Cauchy development of initial data on $\Sigma ,$ c.f. [29, Ch.7],[18].

 By a result of [13] any compact space-like hypersurface is then a Cauchy surface for ({\bf M, g}). It is thus natural to seek a preferred Cauchy surface with respect to which to view the evolution of the space-time. We assume throughout the paper that ({\bf M, g}) has a compact Cauchy surface $\Sigma $ with constant mean curvature
\begin{equation} \label{e0.2}
H = tr K = const <  0. 
\end{equation}
The sign of $H$ is chosen so that $H < $ 0 corresponds to expansion of $\Sigma $ in the future direction. 

 It is a well-known conjecture that any vacuum cosmological space-time admits a CMC Cauchy surface, and this is certainly the case for all known examples; c.f. [10] for further discussion. We will call a cosmological space-time satisfying (0.2) a CMC cosmological space-time.

 One of the reasons for preferring compact CMC surfaces is that any such initial surface $\Sigma $ embeds at least locally, to the past and future, in a smooth foliation ${\cal F}  = \{\Sigma_{\tau}, \tau\in I\}$ of a domain in ({\bf M, g}) by CMC Cauchy surfaces. Each leaf $\Sigma_{\tau}$ is diffeomorphic to $\Sigma $ and the function $H: {\cal F}  \rightarrow  {\Bbb R} $ is monotone decreasing w.r.t. future proper time, c.f. [31]. Let
\begin{equation} \label{e0.3}
M_{{\cal F}} \subset  {\bf M}
\end{equation}
be the maximal domain in {\bf M}
 foliated in this way. Thus, the function
\begin{equation} \label{e0.4}
\tau  = H 
\end{equation}
defines a natural time function on $M_{{\cal F}}.$ We will use $\tau $ to denote the time parameter and $H$ to denote the mean curvature. The range of $\tau : M_{{\cal F}} \rightarrow  {\Bbb R} $ is a connected open interval $I \subset  {\Bbb R} ,$ c.f. [24], and correspondingly, the subset $M_{{\cal F}} \subset $ {\bf M}
 is a domain, i.e. a connected open subset, of {\bf M}.

 The Hawking-Penrose singularity theorem, c.f. [29, Ch.8.2], implies that if ({\bf M, g}) is a CMC cosmological space-time, then there is a singularity in the finite (proper time) past of $\Sigma .$ For the purposes of this Introduction, a singularity to the past (or future) of $\Sigma $ merely means that ({\bf M, g}) is not time-like geodesically complete to the past (or future) of $\Sigma .$ Unfortunately, comparatively little is known in general about the structure of such an "initial" singularity or singularities, c.f. [29], [39].

\medskip

 In this paper, we mainly, (although not only), consider issues concerning the future development of ({\bf M, g}) from the Cauchy surface $\Sigma .$

 Perhaps the first natural issue to consider is to characterize the situation when ({\bf M, g}) is future geodesically complete, i.e. time-like geodesically complete to the future of $\Sigma .$ This depends, at the very minimum, on topological properties of $\Sigma .$ The basic reason derives from one of the constraint equations for the geometry of CMC surfaces in ({\bf M, g}); namely on any leaf $(\Sigma_{\tau}, g_{\tau}) \subset  M_{{\cal F}},$ one has
\begin{equation} \label{e0.5}
s = |K|^{2} -  H^{2}, 
\end{equation}
where $s$ is the scalar curvature of $(\Sigma_{\tau}, g_{\tau})$ and $K$ is the second fundamental form of $(\Sigma_{\tau}, g_{\tau}) \subset $ ({\bf M, g}). Now the scalar curvature plays an important role in understanding the geometry and topology of 3-manifolds, just as it does in dimension 2. Thus, define the Sigma constant $\sigma (\Sigma )$ of a closed oriented 3-manifold $\Sigma $ to be the supremum of the scalar curvatures of unit volume Yamabe metrics on $\Sigma ,$ c.f. [1], [36]. This is a topological invariant, which divides the family of closed 3-manifolds naturally into three classes according to
\begin{equation} \label{e0.6}
\sigma (\Sigma ) <  0, \  \sigma (\Sigma ) = 0, \ \sigma (\Sigma ) >  0. 
\end{equation}
By the solution to the Yamabe problem, $\sigma (\Sigma ) > $ 0 if and only if $\Sigma $ admits a metric of positive scalar curvature. In fact, $\sigma (\Sigma ) > $ 0 if $\Sigma $ admits a non-flat metric with $s \geq $ 0, c.f. [12]. 

 If $M_{{\cal F}}$ contains a (non-flat) leaf $\Sigma_{o}$ with $H =$ 0, then (0.5) implies that $s \geq $ 0, so that $\sigma (\Sigma_{o}) > $ 0. Since $M_{{\cal F}}$ is open, ({\bf M, g}) admits CMC Cauchy surfaces with $H > $ 0 and $H < $ 0 and hence the Hawking-Penrose singularity theorem implies that there exist singularities to the finite proper time future and past of $\Sigma_{o}.$

 This observation, together with the behavior in certain model cases, namely the space-homogeneous Bianchi geometries on $S^{3}$ and Kantowski-Sacks geometries on $S^{2}\times S^{1},$ have led to following conjecture, c.f. [9],[34].

{\bf Recollapse Conjecture.}
 If $\sigma (\Sigma ) >$ 0, then 
$${\bf M}
 = M_{{\cal F}}, $$
and the leaves $\Sigma_{\tau}$ of ${\cal F} $ have mean curvature taking on all values monotonically in $(-\infty , +\infty ).$ The space-time ({\bf M, g}) recollapses in finite proper time, i.e. there is a singularity to the finite proper time future of $\Sigma .$ 

\medskip

 This conjecture basically says that if $\Sigma  \subset $ {\bf M}
 topologically {\it  can}  admit a metric of positive scalar curvature, then under the CMC time evolution, it will evolve in finite proper time to a maximal surface $(\Sigma_{o}, g_{o})$ with $H =$ 0 where it realizes positive scalar curvature, and then recollapses to the future in finite time. Perhaps the strongest general result on this conjecture is the work in [24], which resolves it in case ({\bf M, g}) has a past and a future crushing singularity [20], i.e. there exists a sequence of compact space-like hypersurfaces $P_{i}$ with $H(P_{i}) \rightarrow  -\infty $ uniformly, and similarly a such a sequence $F_{i}$ with $H(F_{i}) \rightarrow  +\infty $ uniformly.

 We mention that the only known examples of 3-manifolds $\Sigma $ with $\sigma (\Sigma ) > $ 0 are spherical space forms $S^{3}/\Gamma , S^{2}\times S^{1},$ and connected sums of such manifolds.

\medskip

 The vast majority of 3-manifolds satisfy $\sigma (\Sigma ) \leq $ 0, i.e. admit no metric of positive scalar curvature. This is the case for instance if $\Sigma $ has a $K(\pi, 1)$ factor in its prime or sphere decomposition, c.f. [26]. For such Cauchy surfaces, the space-time ({\bf M, g}) admits no compact maximal hypersurfaces, and so the CMC time evolution is restricted to a subset of the half-line $(-\infty ,$ 0). For much of the paper, we assume then that
$$\sigma (\Sigma ) \leq  0. $$

 A natural generalization of the recollapse conjecture to all values of $\sigma (\Sigma )$ is the following, raised in [32], c.f. also [20].

\medskip

{\bf Global existence problem in CMC time.}
 The CMC foliation ${\cal F}  = \{\Sigma_{\tau}\}$ in ({\bf M, g}
) exists for all allowable CMC time $\tau ,$ i.e. for all $\tau\in (-\infty ,$ 0) if $\sigma (\Sigma ) \leq $ 0 and for all $\tau\in (-\infty , +\infty )$ if $\sigma (\Sigma ) > $ 0.

\medskip

 This problem arises from a general belief that a CMC foliation should avoid any singularities at the boundary of ({\bf M, g}), c.f. also [31].

 The following result is of some relevance to this problem.
\begin{theorem} \label{t 0.1.}
  Suppose $\sigma (\Sigma ) \leq $ 0. Then either the CMC foliation ${\cal F} $ exists for all allowable CMC time or there exists a sequence of points $x_{i}\in\Sigma_{\tau_{i}},$ with $\tau_{i} \rightarrow  \tau_{o} < $ 0, such that the curvature {\bf R} of ({\bf M, g}) blows up on $\{x_{i}\},$ i.e.
$$|{\bf R}|(x_{i}) \rightarrow  \infty . $$
\end{theorem}

 Here $|{\bf R}|$ is measured in terms of the electric-magnetic decomposition of {\bf R}
 with respect to the future unit normal $T$ of the leaves $\Sigma_{\tau}$ of ${\cal F} ,$ i.e.
\begin{equation} \label{e0.7}
|{\bf R}
|^{2} = |E|^{2} + |H|^{2}, 
\end{equation}
where $E$, $H$ are the symmetric bilinear forms $E(X,Y) = \langle R(X,T)T,Y \rangle$, $H(X,Y)$ = $\langle (*R)(X,T)T,Y \rangle$, c.f. [17,\S 0].

 Theorem 0.1 implies that global existence in CMC time fails, (when $\sigma (\Sigma ) \leq $ 0), only when a sequence of leaves $\Sigma_{\tau_{i}}$ of ${\cal F} $ approaches somewhere a {\it  curvature singularity}  at $\partial{\bf M}
,$ c.f. [21]. This behavior is special to CMC foliations, and is not shared by other geometrically defined foliations, i.e. 3+1 decompositions of {\bf M}. For instance the Gauss or proper time-equidistant foliation from a given space-like hypersurface will typically break down before reaching $\partial{\bf M}
.$ The issue of whether $\partial{\bf M}$ consists of {\it  only}  curvature singularities is closely related to the strong cosmic censorship conjecture, c.f. [18,[39]. 

 The assumption $\sigma (\Sigma ) \leq $ 0 in Theorem 0.1 is not necessary and an analogous result is valid for any value of $\sigma (\Sigma ),$ c.f. Remark 2.6.

 Observe also that Theorem 0.1 gives some information on the 'initial' singularity or singularities to the finite past of $\Sigma .$ Namely, it implies that to the past of $\Sigma ,$ either there exists a curvature singularity of {\bf M}
 or the initial singularity is a crushing singularity.

\medskip

 In contrast to the positive case, when $\sigma (\Sigma ) \leq $ 0 one does not expect, even if one has global existence in CMC time, that in general,
$$M_{{\cal F}} = {\bf M}, $$
i.e. the space-time ({\bf M, g}) may well extend past the CMC foliated region. In this context, assuming global CMC time existence, we prove in Theorem 4.1 that the future boundary of $M_{{\cal F}}$ in {\bf M}, i.e. the boundary
$$\partial_{o}M_{{\cal F}} \subset  {\bf M}, $$
to the future of the initial surface $\Sigma $ in (0.2), is a countable union of smooth, connected, complete maximal hypersurfaces $\Sigma_{o} \subset $ {\bf M}. Each component $\Sigma_{o}$ of $\partial_{o}M_{{\cal F}}$ is non-compact, and so not diffeomorphic to $\Sigma .$ However, any smooth compact domain in $\Sigma_{o}$ does embed in $\Sigma .$ The hypersurface $\partial_{o}M_{{\cal F}}$ is a Cauchy surface for the region 
$${\bf M}^{+} = {\bf M} \setminus M_{{\cal F}} $$
to the future of $M_{{\cal F}}$ in {\bf M}. Of course any causal curve starting in $M_{{\cal F}}$ which enters ${\bf M}
^{+}$ can never reenter $M_{{\cal F}}.$

 In analogy to the Hawking-Penrose singularity theorem, (which only applies to {\it  compact}  CMC hypersurfaces), we have:

\medskip

{\bf Singularity Formation Conjecture.}
 If $\partial_{o}M_{{\cal F}} \neq  \emptyset  $ in ({\bf M, g}), then ({\bf M, g}) possesses a singularity to the finite proper-time future in each component of ${\bf M}^{+}.$

\medskip

 It is natural to conjecture that ${\bf M}
^{+}$ itself has a foliation $M_{{\cal F}}^{+}$ by non-compact CMC surfaces limiting on crushing singularities to the finite future, c.f. [20]; this is discussed in more detail in \S 4. The region $M^{+},$ when non-empty, may be expected to be related to the formation of black holes in ({\bf M, g}).

 The global existence problem in CMC time and the singularity formation conjecture thus imply that ({\bf M, g}) is future geodesically complete if and only if $\sigma (\Sigma ) \leq $ 0 and {\bf M} = $M_{{\cal F}}.$

\medskip

 Next, suppose ({\bf M, g}) is a space-time which is future geodesically complete. It is then natural to consider the possible asymptotic behavior of the space-time as the proper time $t$ approaches infinity. This brings us to the relations with the geometrization of the 3-manifold $\Sigma .$

\begin{definition} \label {d0.2.}
 Let $\Sigma $ be a closed, oriented and connected 3-manifold, with $\sigma (\Sigma ) \leq $ 0.

 A {\sf  weak}  geometrization of $\Sigma $ is a decomposition of $\Sigma ,$
\begin{equation} \label{e0.8}
\Sigma  = H \cup  G, 
\end{equation}
where $H$ is a finite collection of complete connected hyperbolic manifolds of finite volume embedded in $\Sigma $ and $G$ is a finite collection of connected graph manifolds embedded in $\Sigma .$ The union is along a finite collection of embedded tori ${\cal T}  = \cup T_{i}, {\cal T}  = \partial H = \partial G.$

 A {\sf  strong}  geometrization of $\Sigma $  is a weak geometrization as above, for which each torus $T_{i} \in  {\cal T} $ is incompressible in $\Sigma ,$ i.e the inclusion of $T_{i}$ into $\Sigma $ induces an injection of fundamental groups.
\end{definition}

 Of course, it is possible that the collection ${\cal T} $ of tori dividing $H$ and $G$ in (0.8) is empty, in which case weak and strong geometrizations are the same. In such a situation, $\Sigma $ is then either a closed hyperbolic manifold or a closed graph manifold.

 For a strong geometrization, the decomposition (0.8) is unique up to isotopy, c.f. [2], [30], but this is far from being the case for a weak geometrization. Any 3-manifold, for instance $S^{3}$ or $T^{3},$ has numerous knots or links whose complement admits a complete hyperbolic metric of finite volume, and hence gives rise to a weak geometrization.

 We recall that a graph manifold is a union of $S^{1}$ fibrations over surfaces, i.e. Seifert fibered spaces, glued together along toral boundary components. More precisely, a graph manifold $G$ has a decomposition into Seifert fibered spaces $S_{i},$ with $\partial S_{i}$ given by a union of tori $\{T_{j}\}.$ The manifold $G$ is then assembled by glueing (some of) the boundary tori together by toral automorphisms. In case $\partial G = \emptyset  ,$ or $\partial G$ consists of incompressible tori, there exists such a decomposition into Seifert fibered pieces along tori $\{T_{j}\},$ each of which is incompressible in $G$. Such an incompressible decomposition, called the JSJ decomposition, is unique up to isotopy, except in a few, well understood special cases. A typical exceptional case are the $Sol$ manifolds, i.e. $T^{2}$ bundles over $S^{1}$ or over an interval or finite quotients of such bundles. The topology of graph manifolds is completely classified, c.f. [40].

 The Seifert fibered spaces above each admit a geometric structure in the sense of Thurston, i.e. a complete locally homogeneous Riemannian metric, c.f. [38], [37]. Thus if $\Sigma $ has a strong geometrization as above, then $\Sigma $ admits a further decomposition by incompressible tori into domains, each of which has a complete geometric structure. Such a structure is called {\it  the}  geometrization of $\Sigma $ in the sense of Thurston.

 Not all 3-manifolds $\Sigma $ have such a geometric decomposition however. The essential 2-spheres $S^{2} \subset  \Sigma $ are obstructions. This may be related to the fact that $M_{{\cal F}} \neq $ {\bf M}
 in general, and is discussed further in Remark 4.3.

\medskip

 Let $t$: {\bf M} $\rightarrow  {\Bbb R} $ be the proper time to a fixed (initial) CMC Cauchy surface $\Sigma $ as in (0.2). Thus, $t(x)$ is the maximal length of time-like curves from $x$ to $\Sigma .$ The maximum is achieved by a time-like geodesic, since ({\bf M, g}) is globally hyperbolic. Similarly, let $t_{\tau}$ be the proper time of the CMC surface $\Sigma_{\tau} \in  M_{{\cal F}}$ to an initial surface $\Sigma ,$ i.e.
\begin{equation} \label{e0.9}
t_{\tau} =
 t_{max}(\tau ) = max\{t(x): x\in\Sigma_{\tau}\}. 
\end{equation}
Thus $t_{\tau}$ is the usual Lorentzian distance $t_{\tau} = t(\Sigma_{\tau}, \Sigma ).$ If ({\bf M, g}) is future geodesically complete, $t_{\tau} \rightarrow  \infty $ as $\tau  \rightarrow $ 0. In this situation, the volume of the slices $\Sigma_{\tau}$ typically diverges to $+\infty $ as $t_{\tau} \rightarrow  \infty ,$ and so it is natural to consider the rescaled metrics
\begin{equation} \label{e0.10}
\bar g_{\tau} = t_{\tau}^{-2}\cdot  g_{\tau}. 
\end{equation}

 We raise the following:

{\bf Weak Global Asymptotics Problem.}

 Suppose $\Sigma $ is closed, oriented, connected and $\sigma (\Sigma ) \leq $ 0, as above. Suppose further that ({\bf M, g}) is future geodesically complete and $M_{{\cal F}} =$ {\bf M}. Then for any sequence $\tau_{i} \rightarrow $ 0, the slices $(\Sigma_{\tau_{i}}, \bar g_{\tau_{i}})$ have a subsequence asymptotic to a weak geometrization of $\Sigma .$

\medskip

 More precisely, the problem states on the region $H \subset  \Sigma $ from (0.8), the metrics $\bar g_{\tau_{i}}$ in (0.10) converge to the complete hyperbolic metric of finite volume on $H$ while on the region $G \subset  \Sigma ,$ the metrics $\bar g_{\tau_{i}}$ collapse the graph manifold $G$ with bounded curvature in the sense of Cheeger-Gromov to a lower dimensional space.

 We note that although (0.10) is a fixed time rescaling of the metric $g_{\tau}$ for each $\tau  = \tau_{i},$ i.e. the scale factor is independent of any base point in $\Sigma_{\tau},$ the spatial behavior of the metrics $\bar g_{\tau_{i}}$ depends strongly on the choice of base points. Some sequences of base points $x_{i}\in\Sigma_{\tau_{i}}$ give rise to hyperbolic limits, while others give rise to a collapse along a graph manifold structure.

 Of course, one may also formulate the corresponding strong problem, that the slices $(\Sigma_{\tau}, \bar g_{\tau})$ are asymptotic to a strong geometrization, or even the geometrization of $\Sigma ,$ as $\tau  \rightarrow $  0.

\medskip

 Further discussion on more detailed aspects of this problem will be given below; see \S 1 and \S 3 for example for details on the collapse behavior. The {\it  conclusion}  of the strong version of this problem is essentially the same as the conclusions of Conjectures I and II of [1], [2] on the behavior of maximizing sequences of Yamabe metrics or minimizing sequences for the $L^{2}$ norm of scalar curvature on a 3-manifold $\Sigma .$

\medskip

 This problem, as well as the previous conjectures and problems, still appear to be very difficult to resolve. We provide some evidence for its validity in the following result. Suppose ({\bf M, g}) is a CMC cosmological space-time satisfying the following {\it  curvature assumption}: there is a constant $C <  \infty $ such that for $x\in $ ({\bf M, g})
\begin{equation} \label{e0.11}
|{\bf R}|(x) + t(x)|\nabla{\bf R}|(x) \leq  \frac{C}{t^{2}(x)}. 
\end{equation}
Observe that the bound (0.11) is scale-invariant.

 We then have the following concerning the asymptotic behavior of $\Sigma_{\tau}$ as $\tau  \rightarrow $ 0, $(t_{\tau} \rightarrow  \infty ).$

\begin{theorem} \label{t 0.3.}
  Let ({\bf M, g}) be a cosmological CMC space-time, with $\sigma (\Sigma ) \leq $ 0. Suppose that the curvature assumption (0.11) holds, and
$$M_{{\cal F}} = {\bf M}
. $$

 Then ({\bf M, g}) is future geodesically complete and, for any sequence $\tau_{i} \rightarrow $ 0, the slices $(\Sigma_{\tau_{i}}, \bar g_{\tau_{i}})$ have a subsequence asymptotic to a weak geometrization of $\Sigma .$
\end{theorem}

  Theorem 0.3 is proved in \S 3, where we also describe the asymptotic behavior of the space-time ({\bf M, g}), c.f. \S3.1 and \S3.3.

 It is an interesting open problem whether any such weak geometrization is in fact a strong geometrization. This issue is discussed further in \S 3.3 where it is shown to be closely related to the recollapse conjecture. Note that since the weak decomposition (0.8) is not unique, different sequences $\tau_{i} \rightarrow $ 0 may possibly give rise to distinct decompositions (0.8). 

 The bound on the derivative of the curvature {\bf R} in (0.11) is needed for purely technical reasons, related to the use of the stability theorem for the Cauchy initial value problem. Of course one would like to remove this dependence on $\nabla {\bf R}$. In Theorem 3.1, we prove a version of Theorem 0.3 without the bound on $\nabla{\bf R}
,$ but requiring an extra hypothesis on the decay of the mean curvature $H$ as $t_{\tau} \rightarrow  \infty .$ A more difficult problem is whether the decay assumption on {\bf R} in (0.11) is really necessary, see \S 5  for further discussion. 

\medskip

 Of course the basic reason that one may expect that hyperbolic manifolds arise from the long-time behavior of $\Sigma_{\tau}$ is the simple fact that the Lorentzian cone on a hyperbolic 3-manifold, i.e.
\begin{equation} \label{e0.12}
{\bf g}
_{o} = - dt^{2} + t^{2}g_{- 1} 
\end{equation}
is a {\it flat}  Lorentzian space-time, and so in particular a vacuum space-time; here $g_{- 1}$ is any complete metric of curvature $- 1$ on a 3-manifold $\Sigma ,$ so that $\Sigma  = H^{3}(- 1)/\Gamma ,$ where $\Gamma $ is a properly discontinuous subgroup of Isom$(H^{3}(- 1))$. In fact, the flat metric ${\bf g}
_{o}$ in (0.12) is just the quotient of the future light cone about a point \{0\} in flat Minkowski space $({\Bbb R}^{4}, \eta )$ by the action of $\Gamma $ extended to the isometry group of $({\Bbb R}^{4}, \eta ),$ i.e. a L\"obell space-time, c.f. [29, p.274].

 In this respect, it has recently been proved by Andersson-Moncrief [8] that if Cauchy data on a given compact CMC surface $\Sigma $ are sufficiently small perturbation of hyperbolic data as in (0.12), and the hyperbolic metric on $\Sigma $ is rigid among conformally flat metrics on $\Sigma ,$ then the resulting space-time ({\bf M, g}) is future geodesically complete, and asymptotic to the flat Lorentzian metric (0.12) as $\tau  \rightarrow $ 0. This is a compact analogue of the result of Christodoulou-Klainerman [17] on the global stability of Minkowski space.

 We also refer to the interesting work of Fischer-Moncrief [23], where the long-term behavior of a naturally defined Hamiltonian for the space-time evolution is related to the Sigma constant (0.6) of $\Sigma .$

\medskip

  I would like to thank Lars Andersson for interesting and informative discussions on topics related to the paper.

\section{Background and Preliminary Results.}
\setcounter{equation}{0} 

 Let ({\bf M, g})  be a CMC cosmological space-time, with $M_{{\cal F}} \subset $ {\bf  M} as in (0.3), and with time function $\tau  = H$ on $M_{{\cal F}}.$ The 4-metric {\bf g} may then be decomposed into a 3+1 split as
\begin{equation} \label{e1.1}
{\bf g} = -\alpha^{2}d\tau^{2} + g_{\tau}, 
\end{equation}
where $g_{\tau}$ is a Riemannian metric on $\Sigma_{\tau}.$ Hence $g_{\tau}$ may be viewed as a curve of metrics on a fixed slice $\Sigma ,$ by means of diffeomorphisms $\psi_{\tau}: \Sigma  \rightarrow  \Sigma_{\tau}.$ The function $\alpha $ is called the lapse function, and measures the (infinitesimal) time-like spread of the leaves $\Sigma_{\tau},$ since $\alpha^{2} = - {\bf g}(\partial_{\tau}, \partial_{\tau}) > $ 0.

 The vacuum Einstein equations {\bf Ric} = 0 imply, via the Gauss and Gauss-Codazzi equations, constraints on the geometry of $(\Sigma_{\tau}, g_{\tau}) \subset $ ({\bf M, g}). The constraint equations are:
\begin{equation} \label{e1.2}
s = |K|^{2} -  H^{2}, 
\end{equation}
\begin{equation} \label{e1.3}
\delta K = 0. 
\end{equation}
Here $K$ is the $2^{\rm nd}$ fundamental form or extrinsic curvature, given by $K(X,Y) = {\bf g}(\nabla_{X}Y, T)$ where $T$ is the future unit normal, $H = tr_{g}K,$ and $\delta $ is the divergence operator. The sign convention on $K$ agrees with the convention following (0.2). The vacuum Einstein evolution equations are
\begin{equation} \label{e1.4}
{\cal L}_{\partial_{\tau}}g = - 2\alpha K, 
\end{equation}
\begin{equation} \label{e1.5}
{\cal L}_{\partial_{\tau}}K = - D^{2}\alpha  + \alpha (Ric + H\cdot  K -  2K^{2}), 
\end{equation}
where ${\cal L} $ denotes the Lie derivative, $D^{2}$ the Hessian and $K^{2}(X,Y) = \langle K(X), K(Y)\rangle .$

 The choice of time function $\tau  = H$ determines the lapse $\alpha ,$ which satisfies the following lapse equation, derived from the $2^{\rm nd}$ variational formula for volume: 
\begin{equation} \label{e1.6}
-\Delta \alpha  + |A|^{2}\alpha  = \frac{dH}{d\tau} = 1, 
\end{equation}
where $\Delta = tr D^{2}$. The equations (1.2)-(1.6) on the surfaces $(\Sigma_{\tau}, g_{\tau})$ are equivalent to the vacuum Einstein equations on ({\bf M, g}). Finally, the Gauss-Codazzi equations, (in general), give
\begin{equation} \label{e1.7}
dK = {\bf R}
^{T}, 
\end{equation}
i.e. $dK(X,Y,Z) = \nabla_{X}K(Y,Z) -  \nabla_{Y}K(X,Z)$ = {\bf R}$(T,X,Y,Z)$. The operator $d$ in (1.7) is the exterior derivative with respect to the Levi-Civita connection $\nabla $ of $g_{\tau},$ when $K$ is viewed as a 1-form with values in $T\Sigma_{\tau}.$

 The lapse equation (1.6) gives rise to the following elementary and well-known, but important estimates.

\begin{lemma} \label{l 1.1.}
  On any leaf $\Sigma_{\tau},$ the lapse $\alpha $ satisfies the following bounds:
\begin{equation} \label{e1.8}
sup \alpha  \leq  \frac{1}{inf |K|^{2}} \leq  \frac{3}{H^{2}}, 
\end{equation}
and
\begin{equation} \label{e1.9}
inf \alpha  \geq  \frac{1}{sup |K|^{2}} >  0. 
\end{equation}

\end{lemma}
{\bf Proof:}
 This is immediate consequence of the maximum principle applied to the elliptic equation (1.6).

{\endproof}

 Note that $\alpha $ is not scale-invariant - it scales as the square of the distance, i.e. as the metric and so inversely to the norm of the curvature. Namely, if ${\bf g}'  = \lambda^{-2}{\bf g}$  is a rescaling of {\bf g}, then (1.1) becomes
\begin{equation} \label{e1.10}
{\bf g}
'  = -\alpha^{2}\lambda^{-2}d\tau^{2} + \lambda^{-2}g_{\tau}. 
\end{equation}
The space metric $g_{\tau}'  = \lambda^{-2}\cdot  g_{\tau}$ is of course just a rescaling of $g_{\tau}.$ Now $\tau $ is the mean curvature in a given metric and so $\tau'  = \lambda\tau .$ Thus, (1.10) becomes
\begin{equation} \label{e1.11}
{\bf g}'  = - (\lambda^{-2}\alpha )^{2}d(\tau' )^{2} + g_{\tau}', 
\end{equation}
so that
\begin{equation} \label{e1.12}
\alpha'  = \lambda^{-2}\alpha . 
\end{equation}
In particular, under this rescaling, the lapse equation is scale-invariant, as are the other equations (1.2)-(1.7), since the vacuum Einstein equations are scale-invariant.

 It will be useful to also represent the metric w.r.t. the parameter $t_{\tau}$ defined as in (0.9) in place of $\tau .$ Thus, we have
\begin{equation} \label{e1.13}
{\bf g}
 = -\alpha^{2}(\frac{d\tau}{dt_{\tau}})^{2}dt_{\tau}^{2} + g_{\tau}. 
\end{equation}
Here, the lapse factor $\alpha\frac{d\tau}{dt_{\tau}}$ is scale-invariant. In (1.13), we are thus parametrizing the slices $\Sigma_{\tau}$ by the distance function $t_{\tau};$ since $t_{\tau}$ is the maximal proper time between $\Sigma_{\tau}$ and $\Sigma_{\tau_{o}},$ we have
\begin{equation} \label{e1.14}
\alpha\frac{d\tau}{dt_{\tau}} \leq  1, 
\end{equation}
everywhere on $\Sigma_{\tau}.$

 Next we prove a natural comparison result for cosmological CMC space-times which will be important in the proof of Theorem 0.3. It is essentially a standard argument in Lorentzian comparison geometry.
\begin{proposition} \label{p 1.2.}
   Let $\Sigma_{\tau_{o}}$ be any initial compact CMC surface, with $H_{o} = \tau_{o} < $ 0 in ({\bf M, g}). Then
\begin{equation} \label{e1.15}
- H(\Sigma_{\tau})\cdot  t_{\tau} \leq  3(1 -  \frac{\tau}{\tau_{o}}). 
\end{equation}
Equality holds in (1.15) for some $\tau  = \tau_{1} >  \tau_{o}$ if and only if the domain ${\bf M}
_{[\tau_{o},\tau_{1}]} = \cup_{\tau\in [\tau_{o},\tau_{1}]}\Sigma_{\tau} \subset $ ({\bf M, g}) is flat and is isometric to a time-annulus in a flat Lorentzian cone as in (0.12). In particular, all the leaves $\Sigma_{\tau}, \tau\in [\tau_{o}, \tau_{1}]$ are of constant negative curvature.
\end{proposition}
{\bf Proof:}
 Given the initial surface $\Sigma_{\tau_{o}},$ let $S(r) = \{x\in{\bf M}: t(x, \Sigma_{\tau_{o}}) = r\}$ = $t^{-1}(r),$ where $t$ is the Lorentzian distance as preceding (0.9). Let $\theta $ be the mean curvature of $S(r)$, with the sign convention as following (0.2) or (1.3). Since the space-time is vacuum, the Raychaudhuri equation, c.f. [29, \S 4.4], gives
\begin{equation} \label{e1.16}
-\theta'  = -\frac{1}{3}\theta^{2} -  2\sigma^{2} \leq  -\frac{1}{3}\theta^{2}, 
\end{equation}
where $\sigma $ is the shear, the trace-free part of $2^{\rm nd}$ fundamental form of $S(r)$, along this geodesic congruence and $\theta'  = \nabla_{N}\theta ,$ where $N$ is the future directed unit tangent vector to the geodesics $\gamma $ normal to $\{S(r)\}$. Thus by integration, one immediately obtains, as long as $\theta  \leq $ 0,
\begin{equation} \label{e1.17}
-\frac{1}{\theta (s)} -  \frac{1}{3}s \uparrow , 
\end{equation}
i.e. this function is monotone increasing in $s$ along any future directed geodesic $\gamma .$ Hence, for $s \geq $ 0,
\begin{equation} \label{e1.18}
-\frac{1}{\theta (s)} -  \frac{1}{3}s \geq  -\frac{1}{H(\Sigma_{\tau_{o}})}.  
\end{equation}
Now on $\Sigma_{\tau},$ there is a point $x_{\tau}$ realizing $t_{\tau},$ the maximal value of $t$ on $\Sigma_{\tau}.$ Hence $S(t_{\tau})$ lies to the future of $\Sigma_{\tau},$ but touches $\Sigma_{\tau}$ at the point $x_{\tau}.$ By a standard geometric maximum principle, c.f. [7, Thm. 3.6] for example, we then have
\begin{equation} \label{e1.19}
-\theta (x_{\tau}) \geq  - H(\Sigma_{\tau}), 
\end{equation}
again with the sign convention above understood. Substituting this in (1.18), and using the fact that $s = t_{\tau}$ at $x_{\tau}$ gives (1.15). It is obvious that (1.15) holds in any region where $H = \tau  \geq $ 0.

 If equality holds in (1.15) at some $\tau_{1} >  \tau_{o},$ then equality holds in (1.18) for all $\tau\in [\tau_{o}, \tau_{1}].$ Hence by (1.16), $\sigma  =$ 0 everywhere in the time annulus $t^{-1}[\tau_{o},\tau_{1}]$ and so the geodesic spheres $S(r)$ are everywhere umbilic. Further, the inequality in (1.19) must also be an equality, and hence the maximum principle again implies that $\Sigma_{\tau} = S(t_{\tau}),$ so that the surfaces $\Sigma_{\tau}$ are all equidistants. These facts imply that $({\bf M}_{[\tau_{o}, \tau_{1}]}, {\bf g})$ is a Lorentzian cone of the form {\bf g} $= - dt^{2} + t^{2}g_{\tau_{o}},$ where $g_{\tau_{o}}$ is the metric on $\Sigma_{\tau_{o}},$ up to  rescaling. The vacuum equations (0.1) then imply that {\bf g} must be flat and the metrics $g_{\tau}$ on $\Sigma_{\tau}$ are of constant curvature.
{\endproof}

 Observe that (1.15) holds if ({\bf M, g}) satisfies the strong energy condition ${\bf Ric_{g}} \geq 0$ in place of the vacuum equations (0.1). However, the rigidity statement in the second part of Proposition 1.2 requires the vacuum equations.

 Proposition 1.2 of course immediately implies
$$limsup_{t_{\tau}\rightarrow  \infty}t_{\tau}(- H) \leq  3, $$
when {\bf M}
 is geodesically complete to the future of $\Sigma_{\tau_{o}}.$

 Since $H$ is the mean curvature, the first variational formula for volume gives
$$\frac{d}{ds}vol \Sigma_{s} = -\int_{\Sigma_{s}}H\cdot  fdV, $$
where $s$ is any parametrization for the leaves $\Sigma_{\tau}$ and $f > $ 0 is the length of the associated variation field. For example, if $s = \tau ,$ then $f = \alpha $ as in (1.1). Consider the family $\Sigma_{\tau}$ parametrized w.r.t. $t_{\tau},$ as in (1.13). Then $f = \alpha\frac{d\tau}{dt_{\tau}}$ and hence by (1.14),
\begin{equation} \label{e1.20}
\frac{d}{dt_{\tau}}vol \Sigma_{\tau} \leq  -\int_{\Sigma_{\tau}}HdV = - Hvol \Sigma_{\tau} \leq  \frac{3}{t_{\tau}}vol \Sigma_{\tau}. 
\end{equation}

 This discussion gives the following monotonicity result on the volume growth of the leaves $\Sigma_{\tau}.$
\begin{corollary} \label{c 1.3.}
  The volume of the leaves $\Sigma_{\tau}$ in ({\bf M, g}) satisfies
\begin{equation} \label{e1.21}
\frac{vol \Sigma_{\tau}}{t_{\tau}^{3}} \downarrow , 
\end{equation}
i.e. the ratio is monotone non-increasing in $t_{\tau}$ or $\tau .$  The volume ratio is constant in $[\tau_{o}, \tau_{1}]$ if and only if the corresponding time-annulus in ({\bf M, g}) is a flat Lorentzian cone.
\end{corollary}
{\bf Proof:}
 This follows immediately from (1.20) by integration.

{\endproof}

 We will need several results from the Cheeger-Gromov theory on convergence and collapse of Riemannian manifolds, c.f. [14], [15], [4]. For simplicity, we consider only the 3-dimensional case, where the Ricci curvature determines the full curvature tensor.
\begin{proposition} \label{p 1.4. (Convergence/Non-collapse).}

 Let $(D_{i}, g_{i}, x_{i})$ be a sequence of Riemannian metrics on 3-manifolds $D_{i},$ satisfying
\begin{equation} \label{e1.22}
|Ric_{g_{i}}| \leq  \Lambda , diam_{g_{i}}D_{i} \leq  D, dist_{g_{i}}(x_{i}, \partial D_{i}) \geq  d_{o}, 
\end{equation}
and 
\begin{equation} \label{e1.23}
vol_{g_{i}}B_{x_{i}}(1) \geq  v_{o}, 
\end{equation}
for some constants $\Lambda , D <  \infty ,$ and $v_{o}, d_{o} > $ 0. Then for any $\epsilon  > $ 0 with $\epsilon  <  d_{o},$ there are domains $U_{i} = U_{i}(\epsilon ) \subset  D_{i},$ with $dist_{g_{i}}(\partial D_{i}, \partial U_{i}) <  \epsilon ,$ and diffeomorphisms $\psi_{i}$ of $U_{i},$ such that the metrics $\psi_{i}^{*}g_{i}$ have a subsequence converging in the $C^{1,\beta'}$ topology to a limit $C^{1,\beta}$ Riemannian manifold $(U_{\infty}, g_{\infty}, x_{\infty}), x_{\infty} =$ lim $x_{i},$ for any $\beta'  <  \beta  < $ 1. In particular, $U_{\infty}$ is diffeomorphic to $U_{i},$ for $i$ sufficiently large. 
\end{proposition}

\bbgin{proposition} \label{p 1.5. (Collapse).}
  Let $(D_{i}, g_{i}, x_{i})$ be a sequence of Riemannian metrics on 3-manifolds $D_{i},$ satisfying
\begin{equation} \label{e1.24}
|Ric_{g_{i}}| \leq  \Lambda , diam_{g_{i}}D_{i} \leq  D, dist_{g_{i}}(x_{i}, \partial D_{i}) \geq  d_{o}, 
\end{equation}
and 
\begin{equation} \label{e1.25}
vol_{g_{i}}B_{x_{i}}(1) \rightarrow  0, 
\end{equation}
for some constants $\Lambda , D <  \infty $ and $d_{o} > $ 0. Then there are domains $U_{i} \subset  D_{i}$ as above and diffeomorphisms $\psi_{i}$ of $D_{i},$ such that $U_{i}$ is either a Seifert fibered space or a torus bundle over an interval. In both cases, the $g_{i}$ diameter of any fiber F, (necessarily a circle $S^{1}$ or torus $T^{2}),$ goes to 0 as $i \rightarrow  \infty .$ 

 Further, suppose $U_{i}$ is not diffeomorphic to the closed spherical space-form $S^{3}/\Gamma .$ Then for any $i <  \infty $ sufficiently large, $\pi_{1}(F)$ injects in $\pi_{1}(U_{i})$ and there is a finite cover $\bar U_{i}$ of $U_{i}$ such that the sequence $(\bar U_{i}, g_{i}, x_{i})$ does not collapse, i.e. satisfies (1.23), and hence there is a subsequence converging in the $C^{1,\beta'}$ topology as above to a $C^{1,\beta}$ Riemannian manifold $(\bar U_{\infty}, g_{\infty}, x_{\infty}).$ The limit $(\bar U_{\infty}, g_{\infty})$ admits a locally free isometric action by one of the following Lie groups: $S^{1}, S^{1}\times S^{1}, T^{3},$ Nil. 
\end{proposition}

 Both of these results essentially remain valid if $diam_{g_{i}}D_{i} \rightarrow  \infty $ as $i \rightarrow  \infty ,$ but now both behaviors convergence/collapse are possible depending on the choice of base points $x_{i}.$ Thus, suppose $(\Sigma , g_{i})$ are complete Riemannian 3-manifolds with 
\begin{equation} \label{e1.26}
|Ric_{g_{i}}| \leq  \Lambda ,  
\end{equation}
for some constant $\Lambda  <  \infty .$ For a given sequence of base points $x_{i}\in\Sigma ,$ suppose (1.23) holds. Then a subsequence of $(\Sigma_{i}, g_{i}, x_{i})$ converges, modulo diffeomorphisms of $\Sigma_{i},$ to a complete $C^{1,\beta}$ limit Riemannian manifold $(\Sigma_{\infty}, g_{\infty}, x_{\infty}).$ The limit $\Sigma_{\infty}$ embeds weakly in $\Sigma ,$ denoted as
\begin{equation} \label{e1.27}
\Sigma_{\infty} \subset\subset  \Sigma , 
\end{equation}
in the sense that any domain with smooth and compact closure in $\Sigma_{\infty}$ embeds smoothly in $\Sigma .$ (Such a result also holds if $\Sigma  = \Sigma_{i}$ varies with $i$, but we will not need this situation).

 On the other hand, if (1.25) holds at $x_{i} \in \Sigma$, then the sequence collapses uniformly on compact sets, in the sense that $vol_{g_{i}}B_{y_{i}}(1) \rightarrow $ 0, for all $y_{i}\in B_{x_{i}}(R),$ for any fixed $R <  \infty .$ In this case, the collapse may be unwrapped as above and one obtains a complete limit $C^{1,\beta}$ Riemannian manifold $(\bar \Sigma_{\infty}, g_{\infty}, x_{\infty}),$ which admits a locally free isometric action by one of the groups $S^{1}, S^{1}\times S^{1}, T^{3}, Nil$.

 The convergence in Propositions 1.4 and 1.5 above is actually in the weak $L^{2,p}$ topology, for any $p <  \infty $ and the limits are $L^{2,p}$ smooth. By Sobolev embedding, convergence in this topology implies convergence in the $C^{1,\beta}$ topology, for any $\beta  < $ 1.

\begin{remark} \label{r 1.6.(i).}
  {\rm If the bound $|Ric_{g_{i}}| \leq  \Lambda $ in (1.26) is strengthened to a bound on the derivatives of the curvature, i.e.
\begin{equation} \label{e1.28}
|\nabla^{j}Ric_{g_{i}}| \leq  \Lambda_{j} <  \infty , 
\end{equation}
then one obtains convergence to a $L^{j+2,p}$ limit in the weak $L^{j+2,p}$ topology, $p <  \infty .$ In particular, if (1.28) holds for all $j$, then one has $C^{\infty}$ convergence to a $C^{\infty}$ limit.}

{\bf (ii).}
 {\rm By standard results in Riemannian comparison geometry, a lower volume bound as in (1.23) under the curvature bound (1.22) is equivalent to a lower bound on the injectivity radius at $x_{i},$ c.f. [33].

 The collapse situation in Proposition 1.5 corresponds to the formation of families of (arbitrarily) short geodesic loops in $(D_{i}, g_{i})$ within bounded distance to $x_{i}$ and the unwrapping of the collapse corresponds to the unwrapping of these short loops to loops of length about 1 in covering spaces. 

 In the single exceptional case of $S^{3}/\Gamma ,$ the collapse is along inessential loops, as for instance in the Berger collapse, c.f. [14], of $S^{3}$ to $S^{2}$ by shrinking the $S^{1}$ fibers of the Hopf fibration. (Exactly this behavior occurs at the Cauchy horizon of the Taub-NUT metric, c.f. [29, \S 5.8]). In all other cases, the short loops are essential in bounded domains about $x_{i},$ (with diameter depending on the degree of collapse in (1.25)), although they are not necessarily essential globally, i.e. in all of $D_{i}.$

 The dimension of the groups $S^{1}, S^{1}\times S^{1}, T^{3}$ or $Nil$ corresponds to the dimension of this family of short loops. For instance, in the latter two cases of $T^{3}$ or $Nil$, one has a 3-dimensional family of short loops, and so {\it  all}  of $(D_{i}, g_{i})$ is collapsing, and in fact
$$diam_{g_{i}}D_{i} \rightarrow  0; $$
c.f. \S 3.3(I) for an example.}

{\bf (iii).}
 {\rm Define the $C^{1,\beta}$ harmonic radius $r_{h}(x) = r_{h}^{1,\beta}(x)$ of a Riemannian 3-manifold $(\Sigma , g)$ at $x$ to be the largest radius of the geodesic ball about $x$ on which there exists a harmonic coordinate chart in which the metric components are controlled in the $C^{1,\beta}$ norm, i.e. as matrices,
\begin{equation} \label{e1.29}
C^{-1}\delta_{ij} \leq  g_{ij} \leq  C\delta_{ij}, 
\end{equation}
and, for $p$ given by $\beta  =$ 1 $-  \frac{3}{p},$
\begin{equation} \label{e1.30}
r_{h}(x)^{p}||\partial g_{ij}||_{C^{\beta}(B_{x}(r_{h}(x)))}\leq  C. 
\end{equation}
The power of $r_{h}$ in (1.30) is chosen so that $r_{h}$ scales as a radius, i.e. as a distance. The assumptions (1.22)-(1.23) imply a lower bound on $r_{h}, r_{h}(x_{i}) \geq  r_{o},$ where $r_{o}$ depends only on $\Lambda , v_{o},$ (and $d_{o}),$ c.f. [4] and references therein. Similar results hold for the $C^{j+1,\beta}$ harmonic radius, defined analogously to (1.29)-(1.30) when a curvature bound of the form (1.28) is used.}
\end{remark}

\section{Curvature Estimates on CMC Surfaces.}
\setcounter{equation}{0} 

 In this section, we prove Theorem 0.1 and several related results. Since ({\bf M, g}) is smooth, for any given compact subset $\Omega  \subset $ {\bf M}, there is a constant $\Lambda_{o} = \Lambda_{o}(\Omega ) <  \infty $ such that, for all $x\in\Omega ,$
\begin{equation} \label{e2.1}
|{\bf R}
|(x) \leq  \Lambda_{o}, 
\end{equation}
where $|{\bf R}
|$ is measured as in (0.7). Of course, on approach to $\partial${\bf M}, this estimate may no longer hold. In the results to follow, we use the bound (2.1) to control the intrinsic and extrinsic geometry of the leaves $\Sigma_{\tau}$ of $M_{\cal F}$.

\medskip

 First, consider the symmetric bilinear form $K$ as a 1-form on $\Sigma_{\tau}$ with values in $T\Sigma_{\tau},$ as in (1.7). We have the following standard Weitzenbock formula, c.f. [11] for example:
\begin{equation} \label{e2.2}
\delta dK + d\delta K = 2D^{*}DK + {\cal R} (K), 
\end{equation}
on $(\Sigma_{\tau}, g_{\tau}),$ where the curvature term is given by
\begin{equation} \label{e2.3}
{\cal R} (K) = Ric\circ K + K\circ Ric -  2R\circ K, 
\end{equation}
Here $R\circ K$ is the action of the curvature tensor $R$ on symmetric bilinear forms given by: $(R\circ K)(X,Y) = \sum \langle R(X,e_{i})K(e_{i}),Y \rangle$,  $\{e_{i}\}$ a local orthonormal framing for $\Sigma_{\tau}.$ The sign of the curvature tensor is such that $R_{ijji}$ denotes the sectional curvature in the $(ij)$ direction. 

 By the constraint equation (1.3), $\delta K =$ 0 while the Gauss-Codazzi equation gives $dK = {\bf R}^{T}.$ Pairing (2.2) with $K$ thus implies
\begin{equation} \label{e2.4}
\Delta |K|^{2} = 2|DK|^{2} -  \langle \delta{\bf R}
^{T}, K \rangle  + \langle {\cal R} (K), K \rangle , 
\end{equation}

 The key point is to prove that the algebraic curvature term in (2.4) is positive, on the order of $|K|^{4}.$

\bbgin{lemma} \label{l 2.1.}
  On a leaf $\Sigma_{\tau},$ with mean curvature $\tau  =$ H, the curvature term in (2.4) satisfies
\begin{equation} \label{e2.5}
\langle {\cal R} (K), K \rangle  \geq  |K|^{4} -  c(|H||K|^{3} + \Lambda_{o}|K|^{2} + H^{2}|K|^{2}), 
\end{equation}
where $c$ is a numerical constant, independent of ({\bf M, g}) and $\Sigma_{\tau}.$
\end{lemma}
{\bf Proof:}
 Let $\lambda_{i}, i = 1,2,3$, be the eigenvalues of $K$. The Gauss equation gives, for $j \neq  k$,
\begin{equation} \label{e2.6}
R_{jkkj} = {\bf R}_{jkkj} -  \lambda_{j}\lambda_{k}. 
\end{equation}
The minus sign in (2.6), which is crucial in the following, uses the fact that ({\bf M, g}) is Lorentzian; in the case of Riemannian geometry, one has a plus sign, (which would give the wrong sign to the dominant term in (2.5)). We also have, since $\Sigma $ is a 3-manifold, $Ric_{ii} = \frac{s}{2} -  R_{jkkj},$ where $(i,j,k)$ are distinct. Hence
$$\langle Ric\circ K, K \rangle  = \langle K\circ Ric, K \rangle  = Ric_{ii}\cdot \lambda_{i}^{2} = \frac{s}{2}\sum\lambda_{i}^{2} -  \sum R_{jkkj}\lambda_{i}^{2} $$

$$=  \frac{s}{2}|K|^{2} + \sum\lambda_{i}^{2}\lambda_{j}\lambda_{k} -  \sum\lambda_{i}^{2}{\bf R}
_{jkkj}. $$
Observe that $\sum\lambda_{i}^{2}\lambda_{j}\lambda_{k} = H\lambda_{1}\lambda_{2}\lambda_{3} = HdetK \leq  c\cdot |H||K|^{3}$ and $s = |K|^{2} -  H^{2}$ by the constraint equation (1.2).

 Similarly, from the definition, one computes
$$\langle R\circ K, K \rangle  = \sum_{i\neq j}\lambda_{i}\lambda_{j}R_{ijji} = \sum_{i\neq j}\lambda_{i}\lambda_{j}({\bf R}
_{ijji} -  \lambda_{i}\lambda_{j}) = \sum_{i\neq j}\lambda_{i}\lambda_{j}{\bf R}
_{ijji} -  \sum_{i\neq j}\lambda_{i}^{2}\lambda_{j}^{2}. $$
Combining these estimates and using the bound (2.1) gives
$$\langle {\cal R} (K), K \rangle  \geq  |K|^{4} -  c(|H||K|^{3} + \Lambda_{o}|K|^{2} + H^{2}|K|^{2}) + 2\sum_{i\neq j}\lambda_{i}^{2}\lambda_{j}^{2}, $$
which implies (2.5).

{\endproof}

\begin{proposition} \label{p 2.2.}
  Suppose (2.1) holds and let $\Sigma_{\tau}$ be any compact leaf of $M_{{\cal F}}, \Sigma_{\tau} \subset  \Omega ,$ with 
\begin{equation} \label{e2.7}
|H| \leq  H_{o} <  \infty , 
\end{equation}
for some constant $H_{o} > $ 0. Then there is a constant $\Lambda_{1} = \Lambda_{1}(\Lambda_{o}, H_{o})$ such that, on $(\Sigma_{\tau}, g_{\tau}),$
\begin{equation} \label{e2.8}
|K|_{L^{\infty}} \leq  \Lambda_{1}, |Ric|_{L^{\infty}} \leq  \Lambda_{1}. 
\end{equation}
\end{proposition}

{\bf Proof:}
 By the Gauss equation (2.6) and the bound (2.1), the two estimates in (2.8) are equivalent, so it suffices to prove the first estimate.

 Since $\Sigma_{\tau}$ is compact, we may choose a point $x\in\Sigma_{\tau}$ realizing the maximal value of $|K|$ on $\Sigma_{\tau},$ i.e.
\begin{equation} \label{e2.9}
|K|(y) \leq  |K|(x) \equiv  \bar K, 
\end{equation}
for all $y\in\Sigma_{\tau}.$ We will prove (2.8) by contradiction, although with some further work it is possible to give a direct argument, (without passing to the limit below), in which case the dependence of $\Lambda_{1}$ on $(\Lambda_{o}, H_{o})$ is more explicit.

 Thus, if (2.8) is false, there exist compact CMC surfaces $(\Sigma_{i}, g_{i})$ in space-times $({\bf M_{i}, g_{i}})$ satisfying (2.1) and (2.7), but for which $\bar K_{i} = |K_{i}|(x_{i}) = sup_{\Sigma_{i}}|K_{i}| \rightarrow  \infty .$

 Consider the rescaled metric
\begin{equation} \label{e2.10}
\tilde g_{i} = (\bar K_{i})^{2}\cdot  g_{i} 
\end{equation}
on $\Sigma_{i},$ and the corresponding space-time $({\bf M_{i}, \tilde g_{i}}).$ By the scaling properties of the extrinsic curvature $K$, (2.9) becomes
\begin{equation} \label{e2.11}
|\tilde K_{i}|(y_{i}) \leq  |\tilde K_{i}|(x_{i}) = 1 
\end{equation}
on $(\Sigma_{i}, \tilde g_{i}).$ Similarly, the scaling properties of the curvature {\bf R} and mean curvature $H$ imply
\begin{equation} \label{e2.12}
|{\bf \tilde R_{i}}
| \leq  \Lambda_{o}\cdot  (\bar K_{i})^{-2} \rightarrow  0, \ \ {\rm as} \ \ i \rightarrow  \infty , 
\end{equation}
\begin{equation} \label{e2.13}
|\tilde H_{\Sigma_{i}}| \leq  H_{o}(\bar K_{i})^{-1} \rightarrow  0, \ \ {\rm as} \ \ i \rightarrow  \infty , 
\end{equation}
on $({\bf M_{i}, \tilde g_{i}})$ and $(\Sigma_{i}, \tilde g_{i})$ respectively. Hence, by the Gauss equation (2.6), we obtain
\begin{equation} \label{e2.14}
|Ric_{\tilde g_{i}}|(x_{i}) \geq  \frac{1}{2}, \ \ {\rm while} \ \  |Ric_{\tilde g_{i}}|(y_{i}) \leq 2, 
\end{equation}
for all $y_{i}\in (\Sigma_{i}, \tilde g_{i}).$

 Thus, the manifold $(\Sigma_{i}, \tilde g_{i})$ has uniformly bounded curvature. By Proposition 1.5, if the geodesic ball $(\tilde B_{x_{i}}(10), \tilde g_{i})$ in $\Sigma_{i}$ is sufficiently collapsed, we may then unwrap the collapse by passing to a sufficiently large finite cover, so that vol $\tilde B_{x_{i}}(1) \geq  10^{-1}.$ We assume this has been done, without change of the notation.

 It follows from Proposition 1.4 and Remark 1.6(iii) that the local $C^{1,\beta}$ geometry of $(\Sigma_{i}, \tilde g_{i})$ is uniformly controlled, (independent of $(\Sigma_{i}, \tilde g_{i})),$ in the ball $(\tilde B_{x_{i}}(10), \tilde g_{i}).$ Thus, there are harmonic coordinate charts on balls $\tilde B(r_{o})$ of a uniform size $r_{o} > $ 0, within $\tilde B_{x_{i}}(10),$ in which the metric components are uniformly controlled in $C^{1,\beta}$ norm.

 Next, the Gauss-Codazzi equation (1.7) and the constraint equation (1.3) give
\begin{equation} \label{e2.15}
d\tilde K = {\bf \tilde R^{T}}, \ \  \delta\tilde K = 0 
\end{equation}
on $(\Sigma_{i}, \tilde g_{i}),$ where $\tilde K = \tilde K_{i}.$ The system (2.15) is first order uniformly elliptic system for $\tilde K.$ In local harmonic coordinates, the coefficients of this system are uniformly bounded in $C^{1,\beta}.$ Hence, from the bound (2.11) and (2.12), standard elliptic interior regularity estimates imply a uniform bound
\begin{equation} \label{e2.16}
|\tilde K|_{L^{1,p}}  \leq  C, 
\end{equation}
where $p <  \infty , C$ depends only on $p$, and the $L^{1,p}$ norm is taken over any ball $\tilde B(r_{o})$ in $\tilde B_{x_{i}}(5).$ In particular, choosing $p > $ 3 and using Sobolev embedding, (which is applicable again by the uniform $C^{1,\beta}$ local control on the metric),  we may find a uniform $r_{1} > $ 0 such that
\begin{equation} \label{e2.17}
osc_{\tilde B(r_{1})}|\tilde K| \leq  \frac{1}{4}, 
\end{equation}
on all balls $\tilde B(r_{1}) \subset  \tilde B_{x_{i}}(5).$ 

 By Proposition 1.4, the manifolds $(\Sigma_{i}, \tilde g_{i}, x_{i})$ have a subsequence converging in the $C^{1,\beta'}$ topology to a $C^{1,\beta}$ limit $(\Sigma_{\infty}, \tilde g_{\infty}, x_{\infty}).$ The equation (2.2) holds on each $\Sigma_{i}$ and so it holds weakly on the limit, i.e. when paired by integration with any $L^{1,p}$ test form of compact support. Hence, since ${\bf \tilde R}_{i} \rightarrow $ 0 in $L^{\infty}$ by (2.12), on the limit $(\Sigma_{\infty} \tilde g_{\infty}),$ the limit form $K = \tilde K_{\infty}$ satisfies
\begin{equation} \label{e2.18}
2D^{*}DK = {\cal R} (K) 
\end{equation}
weakly in $L^{1,p}.$ This is an elliptic equation with $C^{1,\beta}$ coefficients, so elliptic regularity, c.f. [25, Ch. 9.6], implies that $K$ is $L^{3,p}$ smooth, for any $p < \infty$, and hence by Sobolev embedding, $K$ is $C^{2,\beta}$ smooth. In particular, as in (2.4), we then have
\begin{equation} \label{e2.19}
\Delta |K|^{2} = 2|DK|^{2} + <{\cal R} (K), K> . 
\end{equation}
However, at the limit base point $x_{\infty}, |K|$ is maximal, so $\Delta |K|^{2} \leq $ 0. Further, the estimates (2.12)-(2.13) together with (2.5) imply that the curvature term in (2.19) is non-negative.

 Hence, at $x_{\infty},$ we must have $|K| =$ 0. However, this contradicts the estimates (2.11) and (2.17), which pass to the limit. This contradiction thus establishes (2.8).

{\endproof}

 By Remark 1.6(iii), the estimate (2.8) on the Ricci curvature gives apriori local $C^{1,\beta}$ control on the metric $g_{\tau}$ in harmonic coordinates, i.e. an apriori lower bound on the $C^{1.\beta}$ harmonic radius at any $x\in\Sigma_{\tau},$ 
\begin{equation} \label{e2.20}
r_{h}(x) \geq  r_{o} = r_{o}(\Lambda_{o}, H_{o}) >  0, 
\end{equation}
provided one has a lower bound on the volume $vol_{g_{\tau}}B_{x}(1).$ 

 This will not be the case when $vol_{g_{\tau}}B_{x}(1)$ is too small, but in that case $B_{x}(1)$ may be unwrapped by passing to covering spaces as in Proposition 1.5, (assuming $B_{x}(1) \neq  S^{3}/\Gamma ).$ Thus, one obtains such a lower bound on $r_{h}(x)$ in the covering.

 Given such a lower bound on $r_{h},$ one may apply standard elliptic estimates, c.f. [25, \S 8.8], to the lapse equation (1.6) to control the behavior of $\alpha ,$ since the coefficients of $\Delta  $ are controlled in $C^{1,\beta}$ in harmonic coordinates on $B_{x}(r_{h}(x)).$ (Note that the lapse equation is invariant under coverings). Thus, together with the uniform bound on $|K|$ in (2.8), it follows that
\begin{equation} \label{e2.21}
\frac{sup \alpha}{inf \alpha} \leq  A_{o} <  \infty , 
\end{equation}
where the sup and inf are taken over any $B_{x}(r_{o}) \subset  \Sigma_{\tau},$ and $A_{o}$ depends only on $\Lambda_{o}, H_{o}.$ Similarly, given any fixed $x\in\Sigma_{\tau},$ if $\alpha $ is renormalized so that $\bar \alpha = \alpha /\alpha (x),$ then elliptic regularity [25, \S 9.5] applied to the lapse equation (1.6) divided by $\alpha (x)$ gives a uniform bound
\begin{equation} \label{e2.22}
|D^{2}\bar \alpha|_{L^{p}(B_{x}(r_{o}))} \leq  A_{1}, 
\end{equation}
where $A_{1}$ depends only on $\Lambda_{o}, H_{o};$ the estimate (2.22) is assumed to be taken over the local unwrapping covering in case $\Sigma_{\tau}$ is sufficiently collapsed at $x$. In particular, by Sobolev embedding, this gives
\begin{equation} \label{e2.23}
|\nabla\bar \alpha|_{L^{\infty}} \leq  A_{2}, 
\end{equation}
with $A_{2} = A_{2}(\Lambda_{o}, H_{o}).$

 Similarly, consider the equation (2.2) again, i.e.
$$2D^{*}DK = \delta{\bf R}
^{T}+ {\cal R} (K). $$
This is a second order elliptic system in $K$, and the terms ${\bf R}
^{T}$ and ${\cal R} (K)$ are bounded in $L^{\infty}$ by (2.1) and (2.8). Thus, consider the elliptic operator $D^{*}D$ as a mapping
$$D^{*}D: L^{1,p} \rightarrow  L^{-1,p}. $$
In local harmonic coordinates, i.e. within the harmonic radius, the coefficients of $D^{*}D$ are controlled in $C^{1,\beta}.$ It follows from elliptic regularity and the bounds above that $K$ is controlled in $L^{1,p},$ for any $p <  \infty ,$ i.e.
\begin{equation} \label{e2.24}
||K||_{L^{1,p}(B(r_{o}))} \leq  K_{1}, 
\end{equation}
where $K_{1}$ depends only on $\Lambda_{o}, H_{o}.$ 

\medskip

 There is of course no general apriori lower bound on the volumes vol $B_{x}(1),$ (c.f. also the collapse discussion in \S 3). However, if some initial CMC surface $\Sigma_{\tau_{o}}$ as in (0.2) is given which has a fixed lower bound on vol $B_{p}(1), \forall p\in\Sigma_{\tau_{o}},$ then the following result shows that one has a lower bound on vol $B_{x}(1)$ at all points $x\in\Sigma_{\tau}$ within bounded proper distance to $\Sigma_{\tau_{o}}.$

\bbgin{lemma} \label{l 2.3.}
  Let $\Sigma_{\tau_{o}}$ be a CMC surface in the space-time ({\bf M, g}) satisfying (2.1), and suppose
\begin{equation} \label{e2.25}
vol B_{p}(1) \geq  \nu_{o} >  0, 
\end{equation}
for all $p\in\Sigma_{\tau_{o}}.$ Let $\Sigma_{\tau}$ be any other compact CMC surface in ({\bf M, g}) and suppose, for $x\in\Sigma_{\tau}, t(x) = dist_{{\bf g}}(x, \Sigma_{\tau_{o}}) \leq  T_{o}.$ Then there is a constant $\nu_{1} = \nu_{1}(H_{o}, \Lambda_{o}, \nu_{o}, T_{o}) > $ 0, such that
\begin{equation} \label{e2.26}
vol B_{x}(1) \geq  \nu_{1}. 
\end{equation}
In particular, there is then a constant $r_{1} = r_{1}(H_{o}, \Lambda_{o}, \nu_{o}, T_{o}) > $ 0 such that
\begin{equation} \label{e2.27}
r_{h}^{1,\beta}(x) \geq  r_{1} >  0. 
\end{equation}
\end{lemma}

{\bf Proof:}
 Given $x\in\Sigma_{\tau},$ we consider neighborhoods $U$ of $x$ in $\Sigma_{\tau}$ as graphs over $\Sigma_{\tau_{o}}$ via the time-like normal exponential map from $\Sigma_{\tau_{o}}.$ Thus, if $p\in\Sigma_{\tau_{o}}$ is such that $dist_{{\bf g}}(x, p) = dist_{{\bf g}}(x, \Sigma_{\tau_{o}}),$ let $U$ be the set of points $y$ in $\Sigma_{\tau}$ for which there exists a maximal geodesic from $y$ to $B_{p}(1)\subset\Sigma_{\tau_{o}}$ realizing $dist_{{\bf g}}(y, \Sigma_{\tau_{o}}).$ Hence the normal exponential map to $\Sigma_{\tau_{o}}$ induces a continuous map $F: B_{p}(1) \rightarrow  U, F(q) = exp(\phi (q)\cdot  T),$ where $T$ is the time-like unit normal to $\Sigma_{\tau_{o}}$ and $\phi $ is a positive (or negative) function.

 We claim that $F$ is a quasi-isometry, with distortion factor depending only on the bounds $(H_{o}, \Lambda_{o}, T_{o}).$ To see this, consider the 1-parameter interpolation $F_{s}(q) = exp(s\phi (q)\cdot  T).$ The apriori bounds on $|K|$ from (2.8), which controls the infinitesimal distortion of $g_{\tau}$ in the unit normal direction, and $\alpha $ from (2.21), imply that for $s$ small, $F_{s}$ has metric distortion at most $C(H_{o}, \Lambda_{o})\cdot  s$ onto its image. Iterating this control inductively to $s =$ 1 then gives the claim.

 Since $F$ is a quasi-isometry, it is clear that the estimate (2.25) implies (2.26). The bound (2.27) then follows from the arguments preceding the Lemma.

{\endproof}

 Observe that if $H = \tau $ is bounded away from 0, i.e. $|\tau| \geq  \bar \tau > $ 0, then the estimate (1.8) implies that $\alpha $ is bounded above, depending only on $\bar \tau.$ Hence, one has a global estimate for the proper time
\begin{equation} \label{e2.28}
t(x) \leq  T_{o}(\bar \tau), 
\end{equation}
$\forall x \in \Sigma_{\tau}$ in this case.

 We need one further apriori estimate for the proof of Theorem 0.1.

\bbgin{lemma} \label{l 2.4.}
  For $\Sigma_{\tau} \subset  \Omega ,$ as in (2.1), there is a constant $\Lambda_{1} <  \infty ,$ depending only on $H_{o}, \Lambda_{o},$ vol $\Sigma_{\tau}$ and an initial surface $\Sigma_{\tau_{o}}$ as in (0.2) such that
\begin{equation} \label{e2.29}
||\nabla Ric_{\Sigma_{\tau}}||_{L^{2}} \leq  \Lambda_{1}, \ {\rm and} \ ||\nabla^{2}K_{\Sigma_{\tau}}||_{L^{2}} \leq  \Lambda_{1}. 
\end{equation}
\end{lemma}

{\bf Proof:}
 Given the estimates above from Proposition 2.2 through Lemma 2.3, (2.29) follows from estimates for the $1^{\rm st}$ order Bel-Robinson energy and we refer to [17], [6], [8] for some further details.

 Thus, let $Q^{1}$ be the $1^{\rm st}$ order Bel-Robinson tensor associated to the Weyl-type field ${\bf W}^{1} = \nabla_{T}{\bf W},$ where {\bf W} is the Weyl tensor of ({\bf M, g}) and $T$ is the unit normal to the foliation ${\cal F}.$ Then there are numerical constants $c, C$ such that, for 
\begin{equation} \label{e2.30}
{\cal Q}^{1}(\tau ) =\int_{\Sigma_{\tau}}Q^{1}(T,T,T,T) 
\end{equation}
one has
\begin{equation} \label{e2.31}
c{\cal Q}^{1}(\tau ) \leq  ||\nabla Ric_{\Sigma_{\tau}}||_{L^{2}}^{2} + ||\nabla^{2}K_{\Sigma_{\tau}}||_{L^{2}}^{2} \leq  C{\cal Q}^{1}(\tau ). 
\end{equation}
Further, ${\cal Q}^{1}$ obeys the following differential inequality:
\begin{equation} \label{e2.32}
|\frac{d}{d\tau}{\cal Q}^{1}(\tau )| \leq  M(1 + {\cal Q}^{1}(\tau ) + F(\tau )), 
\end{equation}
where $M$ depends only on the $L^{\infty}$ norm of $K$ and $|\nabla log \alpha|$ on $\Sigma_{\tau}$ and
$$F(\tau ) = \int_{\Sigma_{\tau}}(|Ric|^{4} + |DK|^{4}). $$
By (2.8) and (2.23), $M$ is uniformly bounded in terms of $\Lambda_{o}, H_{o}$ while (2.8) and (2.1) give, via the H\"older inequality, a uniform bound on $F(\tau )$ depending only on $\Lambda_{o}, H_{o}$ and an upper bound for $vol \Sigma_{\tau}.$ This gives uniform control on $M$ and $F(\tau )$ on the right side of (2.32), and hence (2.29) follows by integration w.r.t. $\tau .$

{\endproof}

\bbgin{remark} \label{r 2.5.}
  {\rm In \S 4, we will also use the analogous $0^{\rm th}$ order Bel-Robinson estimate. Thus, the Bel-Robinson energy ${\cal Q}^{o}(\tau ),$ given by
\begin{equation} \label{e2.33}
{\cal Q}^{o}(\tau ) = \int_{\Sigma_{\tau}}|Ric|^{2} + |dK|^{2}, 
\end{equation}
satisfies
\begin{equation} \label{e2.34}
|\frac{d}{d\tau}{\cal Q}^{o}(\tau )| \leq  c \cdot M{\cal Q}^{o}(\tau ), 
\end{equation}
where $M = |K|_{L^{\infty}} + |\nabla log\alpha|_{L^{\infty}}$ on $\Sigma_{\tau},$} and $c$ is a numerical constant.
\end{remark}

 This discussion now easily leads to the proof of Theorem 0.1.

\medskip

{\bf Proof of Theorem 0.1.}

 Suppose $\sigma (\Sigma ) \leq $ 0 and let $\Sigma_{\tau_{i}}$ be a sequence of compact leaves in $M_{{\cal F}},$ with $\tau_{i} \rightarrow  \bar \tau < $ 0. We may assume that $\{\tau_{i}\}$ is either increasing or decreasing. Suppose $\Sigma_{\tau_{i}}$ does not approach a curvature singularity of ({\bf M, g}), i.e. the estimate (2.1) holds, for some constant $\Lambda_{o}$ uniformly on $\{\Sigma_{\tau_{i}}\}.$ 

 Since $\bar \tau < $ 0, the estimate (2.28) implies that the proper time function $t$ is uniformly bounded above on $\{\Sigma_{\tau_{i}}\}.$ Thus, the proof of Lemma 2.3 implies that the surfaces $\{\Sigma_{\tau_{i}}\}$ are all uniformly quasi-isometric. Together with the bound on $|Ric|$ from (2.8), we see that the hypotheses (1.22)-(1.23) of Proposition 1.4 hold on $\{\Sigma_{\tau_{i}}\}.$

 Proposition 1.4, Remark 1.6(i) and Lemma 2.4 thus imply that a subsequence of $\{\Sigma_{\tau_{i}}\}$ converges in the weak $L^{3,2}$ topology to a limit $L^{3,2}$ CMC surface $\Sigma_{\bar \tau} \subset $ ({\bf M, g}) with $L^{2,2}$ extrinsic curvature $K_{\bar \tau}.$ Similarly, the extrinsic curvatures $K_{i}$ of $\Sigma_{\tau_{i}}$ converge in weak $L^{2,2}$ to the limit $K_{\bar \tau}$ on $\Sigma_{\bar \tau}.$ Now the Cauchy problem on $\Sigma_{\bar \tau}$ with data $(g_{\bar \tau}, K_{\bar \tau})$ in $(L^{3,2}, L^{2,2})$ is uniquely locally solvable, c.f. [22], and hence there exist unique local developments of this data for a short time into the past and future of $\Sigma_{\bar \tau}.$ By the uniqueness, such a development must lie in a region of the $C^{\infty}$ smooth space-time ({\bf M, g}). Hence, by elliptic regularity applied to the equation (2.2), it follows that $\Sigma_{\bar \tau}$ is $C^{\infty}$ smooth, with $C^{\infty}$ extrinsic curvature. The limit CMC surface $\Sigma_{\bar \tau}$ is the unique surface in ({\bf M, g}) with mean curvature $\bar \tau,$ [31], and hence the sequence $\{\Sigma_{\tau_{i}}\},$ (and not just a subsequence), converges smoothly to the limit $\Sigma_{\bar \tau}.$ Thus, the smooth foliation $M_{{\cal F}}$ extends a definite amount past $\Sigma_{\bar \tau}.$ 

 This completes the proof.

{\endproof}

\begin{remark} \label{r 2.6.}
  {\rm The only place the assumption $\sigma (\Sigma ) \leq $ 0 was used in the proof of Theorem 0.1 was to obtain a uniform upper bound on the lapse function $\alpha $ from (1.8), which in turn was used only to obtain an upper bound on the proper time to $\{\Sigma_{\tau_{i}}\}$ via (2.28). Hence, Theorem 0.1 remains true for any value of $\sigma (\Sigma )$ whenever there is an upper bound for $t_{\tau}$ on $\{\Sigma_{\tau_{i}}\}.$ More generally, since Lemma 2.3 and the results preceeding it are all local, one needs only an upper bound on $t(x)$ on domains in $\{\Sigma_{\tau_{i}}\}.$ This will be discussed further in \S 4.}
\end{remark}

\bbgin{remark} \label{r 2.7.}
  {\rm As remarked in the Introduction, Theorem 0.1 may be applied to the past of the initial surface $\Sigma ,$ for any value of $\sigma (\Sigma )$ by Remark 2.6. It implies that either there is a curvature singularity in the finite proper past of $\Sigma ,$ or one has global CMC time existence to the past, and in the limit $\tau  \rightarrow  -\infty ,$ the space-time ({\bf M, g}) approaches a crushing singularity.

 Observe that the globally hyperbolic space-time ({\bf M, g}) does not extend, as a globally hyperbolic space-time, to the past of a (compact) crushing singularity. To see this, suppose $({\bf M' , g'})$ is such a globally hyperbolic extension of ({\bf M, g}), and choose a point $x\in  {\bf M'}\setminus {\bf M}.$ Since ${\bf M'}$ is globally hyperbolic, there is a maximal time-like geodesic from $x$ to any leaf $\Sigma_{\tau},$ which realizes the maximal distance of $x$ to $\Sigma_{\tau}.$ Hence $\gamma $ has no focal points. But the standard argument in the Hawking singularity theorem via the Raychaudhuri equation (1.16), implies that $\gamma $ must have a focal point if $\tau $ is chosen sufficiently large and negative.

 It follows that in any extension $({\bf M' , g'})$ of ({\bf M, g}), the boundary $\partial_{P}M_{{\cal F}}$ of $M_{{\cal F}}$ to the past of $\Sigma_{\tau_{o}}$ is the past Cauchy horizon of $\Sigma .$ Examples with this behavior of course do occur, most notably in the  Taub-NUT metric, c.f. [29, Ch.5.8].}
\end{remark}

\section{Asymptotics of Future Complete Space-Times.}
\setcounter{equation}{0}

 This section is mainly concerned with the proof of Theorem 0.3. In \S 3.1 we prove the result under an extra hypothesis concerning the asymptotic behavior of the mean curvature $H = \tau ,$ but without the bound on $\nabla{\bf R}
$ in (0.11). This hypothesis is then removed, i.e. proved to always hold, in \S 3.2. In \S 3.3 we make a number of remarks on the collapse situation.

\medskip

 First however we note that a space-time satisfying the assumptions of Theorem 0.3 must be geodesically future complete. In fact, the much weaker assumption of a uniform curvature bound
\begin{equation} \label{e3.1}
|{\bf R}| \leq  \Lambda_{o} <  \infty  
\end{equation}
to the future of an initial surface $\Sigma_{\tau_{o}}$ implies by Theorem 0.1 that the CMC foliation ${\cal F} $ exists for all CMC time $\tau\in [\tau_{o},$ 0). Hence, since by the assumption in Theorem 0.3 that $\partial_{o}M_{{\cal F}} = \emptyset  $ in {\bf M}, it follows that ({\bf M, g}) is future geodesically complete. 

 The remainder of the proof is thus concerned only with the weak geometrization of $\Sigma .$

\medskip

{\bf \S 3.1.}
 As mentioned above, we first prove Theorem 0.3 in a special case. Thus, in contrast to (0.9), define
\begin{equation} \label{e3.2}
t_{min}(\tau ) = min\{t(x): x\in\Sigma_{\tau}\}. 
\end{equation}
As in (1.15), the product $H\cdot  t_{min}$ is scale invariant.

\bbgin{theorem} \label{t 3.1.}
  Suppose ({\bf M, g}) is a CMC cosmological space-time with {\bf M} = $M_{{\cal F}},$ (to the future of $\Sigma_{\tau_{o}}),$ and that ({\bf M, g}) satisfies the curvature assumption
\begin{equation} \label{e3.3}
|{\bf R}
|(x) \leq  \frac{C}{t^{2}(x)}, 
\end{equation}
for some $C <  \infty ,$ compare with (0.11). Further, given a sequence $\{\tau_{i}\} \rightarrow $ 0, suppose there exists a constant $\delta  > $ 0 such that
\begin{equation} \label{e3.4}
|H_{\Sigma_{\tau_{i}}}| \geq  \frac{\delta}{t_{min}(\tau_{i})}. 
\end{equation}

 Then there is a subsequence of $\{\tau_{i}\},$ denoted also $\tau_{i},$ such that the rescaled metrics $(\Sigma_{\tau_{i}}, \bar g_{\tau_{i}})$ as in (0.10) converge to a weak geometrization of $\Sigma .$
\end{theorem}

{\bf Proof:}
 Given a sequence of surfaces $\Sigma_{\tau_{i}}, \tau_{i} \rightarrow $ 0, to simplify notation, we let $\Sigma_{i} = \Sigma_{\tau_{i}}.$ Then combining the assumption (3.2) with the general bound (1.15) gives
\begin{equation} \label{e3.5}
\frac{\delta}{t_{min}(\tau_{i})} \leq  |H_{\Sigma_{i}}| \leq  \frac{3}{t_{max}(\tau_{i})} = \frac{3}{t_{\tau_{i}}}, 
\end{equation}
and hence on each $\Sigma_{i},$
\begin{equation} \label{e3.6}
t_{\tau_{i}} = t_{max}(\tau_{i}) \leq  \frac{3}{\delta}t_{min}(\tau_{i}). 
\end{equation}
This means that the surfaces $\Sigma_{i}$ lie in time-annuli of uniformly bounded ratios, i.e.
\begin{equation} \label{e3.7}
\Sigma_{i} \subset  A(\frac{\delta}{3}t_{\tau_{i}}, t_{\tau_{i}}), 
\end{equation}
where $A(r, s) = t_{\tau}^{-1}(r, s)$.

 Consider then the rescaled space-time metric
\begin{equation} \label{e3.8}
{\bf \bar g}
_{i} = t_{\tau_{i}}^{-2}\cdot {\bf g}
, 
\end{equation}
and the corresponding $3+1$ decomposition
\begin{equation} \label{e3.9}
{\bf \bar g}_{i} = -\bar \alpha_{i}^{2}d\bar \tau_{i}^{2} + \bar g_{\bar \tau_{i}}. 
\end{equation}
Note that the vacuum Einstein equations are of course invariant under rescaling. Here, the mean curvature parameter $\tau $ is replaced by $\bar \tau_{i} = \tau\cdot  t_{\tau_{i}}$ and $\bar \alpha_{i} = \alpha /t_{\tau_{i}}.$ Hence the surfaces $\Sigma_{i},$ now denoted by $\bar \Sigma_{i}$ in this rescaled metric, have mean curvature $\bar H_{i} = \bar H(\bar \Sigma_{i})$ satisfying
\begin{equation} \label{e3.10}
\delta  \leq  |\bar H_{i}| \leq  3. 
\end{equation}
Let $\bar t_{i}$ be the proper time function w.r.t. ${\bf \bar g}
_{i},$ so that $\bar t_{i} = t/t_{\tau_{i}}.$ Thus, (3.7) translates into the statement that
\begin{equation} \label{e3.11}
\bar \Sigma_{i} \subset  \bar A(\frac{\delta}{3}, 1), 
\end{equation}
where $\bar A(r, s) = \bar t_{i}^{-1}(r, s)$ so that the surfaces $\bar \Sigma_{i}$ lie in time-annuli of uniformly bounded inner and outer radii. Of course, the leaves $\Sigma_{\tau}$ of the foliation ${\cal F} $ are now considered as leaves $(\bar \Sigma_{\bar \tau_{i}}, \bar g_{\bar \tau_{i}})$ of the foliation $\bar {\cal F}_{i}$ of the rescaled space-time $({\bf \bar M}_{i}, {\bf \bar g}_{i}).$

 By scaling properties of curvature and the curvature assumption (3.3), we have
\begin{equation} \label{e3.12}
|{\bf \bar R}
_{i}|(x) = t_{\tau}^{2}|{\bf R}
|(x) \leq  C\cdot  t_{\tau}^{2}\cdot  t(x)^{-2}, 
\end{equation}
and hence, for any given $\kappa  > $ 0 we have
\begin{equation} \label{e3.13}
|{\bf \bar R}
_{i}| \leq  C = C(\kappa ), 
\end{equation}
within any time-annulus $\bar A(\kappa , \kappa^{-1}).$ Hence, by Proposition 2.2 the leaves $(\bar \Sigma_{\bar \tau_{i}}, \bar g_{\bar \tau_{i}})$ of $\bar {\cal F}_{i}$ within $\bar A(\kappa , \kappa^{-1})$ have uniformly bounded intrinsic curvature and extrinsic curvature $\bar K.$ Let $\bar H$ also denote the mean curvature of the leaves $\bar \Sigma_{\bar \tau_{i}},$ and recall that $\bar H$ is monotone. Since $|\bar K|^{2} \geq  \frac{1}{3}|\bar H|^{2}$ and $|\bar H|$ is bounded away from 0 on $\bar \Sigma_{i}$ by (3.10),  we thus obtain
\begin{equation} \label{e3.14}
c \leq  |\bar K| \leq  c^{-1} 
\end{equation}
for some constant $c = c(\kappa ) > $ 0 in $\bar A(\kappa , \kappa^{-1}).$

 The bound (3.14) applied to the lapse estimates (1.8)-(1.9) in the $\bar g_{i}$ scale immediately gives uniform bounds for the lapse function $\bar \alpha_{i}$ on the leaves $\bar \Sigma_{\bar \tau_{i}}$ within $\bar A(\kappa , \kappa^{-1}),$ i.e.
\begin{equation} \label{e3.15}
1 \leq  \frac{sup \bar \alpha_{i}}{inf \bar \alpha_{i}} \leq  C, 
\end{equation}
where sup and inf are taken over $\bar \Sigma_{\bar \tau_{i}}$ and $C = C(\kappa ).$ At least when (3.4) holds for all $\tau \geq -1$, (3.15) implies that the original lapse function $\alpha $ on ({\bf M, g}) satisfies $\alpha  \sim  t^{2},$ i.e. the ratio $\alpha /t^{2}$ is bounded above and below away from 0 within $A(\kappa t_{\tau_{i}}, \kappa^{-1}t_{\tau_{i}})$ as $t_{\tau_{i}} \rightarrow  \infty .$

 The volume estimate (1.21) is also scale-invariant, and hence we also have a uniform upper bound
\begin{equation} \label{e3.16}
vol_{\bar g_{\tau_{i}}}\bar \Sigma_{i} \leq  V_{o}, 
\end{equation}
for some $V_{o} <  \infty .$ Note however that the diameter of $(\bar \Sigma_{i}, \bar g_{\tau_{i}})$ may or may not be uniformly bounded. The bounds (3.15) on the lapse function do not control the size of the diameter of $(\Sigma_{i}, \bar g_{\tau_{i}}).$ Given the lapse bounds (3.15), the growth of the diameter is determined by the extrinsic curvature, in particular by the growth of its eigenvalues, which are not controlled by bounds on the lapse function. (We remark this is in contrast to the Riemannian situation, where one can obtain apriori diameter bounds, in addition to volume bounds, from (0.1), or just the assumption of non-negative Ricci curvature).

 Thus, in passing to limits, one must choose base points and the form of the limits will then depend on the choice of the base points. Again however note that all base points on $\bar \Sigma_{i}$ lie in fixed time annuli as in (3.11).

 Thus, let $x_{i}\in\bar \Sigma_{i}$ be any sequence of base points and consider the behavior of the pointed sequence $(\bar \Sigma_{i}, \bar g_{\tau_{i}}, x_{i}).$ By Propositions 1.4 and 1.5, we have two possible cases.

\medskip

{\bf Case A.}
 Suppose the sequence $(\bar \Sigma_{i}, \bar g_{\tau_{i}}, x_{i})$ is non-collapsing, in that (1.23) holds with $\bar g_{\tau_{i}}$ in place of $g_{i}.$ This lower volume bound implies that there exists a constant $\nu_{o} > $ 0 such that
\begin{equation} \label{e3.17}
vol_{\bar g_{\tau_{i}}}\bar \Sigma_{i}= \frac{vol \Sigma_{i}}{t_{\tau_{i}}^{3}} \geq  \nu_{o}, 
\end{equation}
for $t_{\tau_{i}} \rightarrow  \infty ;$ the first equality follows from the scale invariance of the ratio. By the monotonicity property (1.21), it follows that (3.17) holds for all $t_{\tau},$ and we may assume that $\nu_{o}$ is the limiting minimal value.

 As noted above, all leaves $\bar \Sigma_{\bar \tau_{i}}$ in annuli $\bar A(\kappa , \kappa^{-1})$ have uniform $L^{\infty}$ bounds on the intrinsic curvature $Ric$ and hence the spatial metrics $\bar g_{\bar \tau_{i}}$ are uniformly controlled locally in $L_{x}^{2,p},$ where $L_{x}^{2,p}$ denotes the spatial Sobolev space, along the leaves $\bar \Sigma_{\bar \tau_{i}}.$ By (2.24), there is a uniform local $L_{x}^{1,p}$ bound on the extrinsic curvature $K$ while by (2.22), the lapse function $\bar \alpha_{i}$ also is uniformly bounded locally in $L_{x}^{2,p}$ in $\bar A(\kappa , \kappa^{-1}).$ 

 Similarly, we claim that the time derivative $\partial_{\tau}^{2}\bar g_{i}$ is uniformly bounded locally in $L^{p}$ in $\bar A(\kappa , \kappa^{-1}),$ where $\tau  = \bar \tau_{i}$ is the time parameter. For by (1.4) and the estimates above, $\partial_{\tau}\bar g_{i} = - 2\bar \alpha_{i}\bar K_{i}$ is uniformly bounded in $L_{x}^{1,p}.$ We have $\partial_{\tau}^{2}\bar g_{i} = - 2(\partial_{\tau}\bar \alpha_{i})\bar K_{i} - 2\bar \alpha_{i}\partial_{\tau}\bar K_{i}.$ The second term is uniformly bounded locally in $L^{p}$ by the estimates above applied to the evolution equation (1.5). Further, by differentiating the lapse equation (1.6) w.r.t. $\tau $ and using standard formulas for the derivative of $\Delta  ,$ (c.f. [12, 1.184]), a straightforward calculation shows $(-\Delta   + |K|^{2})\partial_{\tau}\bar \alpha_{i}$ is uniformly bounded locally in $L_{x}^{p}$ and hence by elliptic regularity, $\partial_{\tau}\bar \alpha_{i}$ is uniformly bounded locally in $L_{x}^{2,p}.$ This proves the claim.

 It follows that the sequence of Lorentzian vacuum space-times in (3.9) has a subsequence converging in the weak $L^{2,p}$ topology to a limit $C^{1,\beta}\cap L^{2,p}$ Lorenztian vacuum space-time, of the form
\begin{equation} \label{e3.18}
{\bf \bar g}_{\infty} = -\bar \alpha_{\infty}^{2}d\bar \tau_{\infty}^{2} + \bar g_{\bar \tau_{\infty}}. 
\end{equation}
(More precisely, one should first take a limit within the time-annuli $\bar A(\kappa , \kappa^{-1})$ and then let $\kappa = \kappa_{j}, j = j(i) \rightarrow $ 0 in a suitable diagonal subsequence). Here $\bar \tau_{\infty}$ is the mean curvature parametrization given by the limit of the parametrizations $\bar \tau_{i}$ from (3.9) and (3.18) is a weak solution of the vacuum equations (0.1).

 Now by the volume monotonicity (1.21) used above, this limit space-time is a 'volume cone' in the sense that
\begin{equation} \label{e3.19}
vol \bar \Sigma_{\bar \tau_{\infty}}/(\bar t_{\infty})^{3} \equiv  \nu_{o} > 0,\end{equation}
where $\bar t_{\infty}$ is the limit of the renormalized proper time functions $\bar t_{i}$ following (3.10). By Corollary 1.3, it thus follows that ${\bf \bar g}_{\infty}$ is flat, and $\bar g_{\bar \tau_{\infty}}$ is complete and hyperbolic, (i.e. of constant negative curvature). Thus, the limit is a flat Lorentzian cone on a hyperbolic manifold, as in (0.12). 

  Note that the limit surface $\bar \Sigma_{\infty} =$ lim $\bar \Sigma_{i}$ may be compact, in which case it is diffeomorphic to $\Sigma ,$ or non-compact. The limit lapse function $\bar \alpha_{\infty} \equiv $ 1, so that the leaves $\bar \Sigma_{\bar \tau_{\infty}}$ are level sets of $\bar t_{\infty}$, while the limit extrinsic curvature $\bar K_{\infty}$ is pure trace, with $|\bar K_{\infty}|^{2} =$ 3 on the limit $\bar \Sigma_{\infty}.$

 Observe that consequently, not only does (3.4) hold, but in fact
\begin{equation} \label{e3.20}
H(\Sigma_{i})\cdot  t(y_{i}) \rightarrow  3, 
\end{equation}
for all $y_{i} \in  \Sigma_{i}$ such that $dist_{\bar g_{\bar \tau_{i}}}(x_{i}, y_{i}) \leq  D$, for any fixed $D <  \infty .$

 Thus, we see that the {\it  only}  limit geometry arising in the non-collapse situation is the hyperbolic geometry.

\medskip

{\bf Case B.}
 Suppose the sequence $(\bar \Sigma_{i}, \bar g_{\tau_{i}}, x_{i})$ is collapsing, so that (1.25) holds, again with $\bar g_{\bar \tau_{i}}$ in place of $g_{i}.$ By (3.12), it is then collapsing with bounded curvature, and thus, for any $R <  \infty ,$ the geodesic $R$-balls about $x_{i}$ in $(\bar \Sigma_{i}, \bar g_{\bar \tau_{i}})$ are Seifert fibered spaces or torus bundles over an interval $I$, with collapsing fibers. In particular, this part of $\bar \Sigma_{i}$ is an (elementary) graph manifold, and corresponds to a piece in the decomposition of the graph manifold $G$, c.f. the discussion in \S 0. 

 As discussed following Proposition 1.5, we may choose a sequence $R = R_{j} \rightarrow  \infty $ and pass to a suitable diagonal subsequence, which, after unwrapping the collapse, converges in $C^{1,\beta}\cap L^{2,p}$ to a complete limit Riemannian manifold $(\bar \Sigma_{\infty}, \bar g_{\infty}, x_{\infty}).$ Note that we use here the assumption $\sigma (\Sigma ) \leq $ 0, so that $\Sigma  \neq  S^{3}/\Gamma .$

 The limit is either a Seifert fibered space or torus bundle over an interval, (i.e. a $Sol$ manifold), and has either a locally free isometric $S^{1}, S^{1}\times S^{1}, T^{3}$ or $Nil$ action. For the same reasons as above in Case A, preceding (3.18), the unwrapping of the collapse also gives rise to a limit space-time $({\bf \bar M}_{\infty}, {\bf \bar g}_{\infty}, x_{\infty})$ and a corresponding maximal CMC foliation $\bar M_{\bar {\cal F_{\infty}}}$ with $\bar \Sigma_{\infty}$ as a leaf. Thus the limit metric ${\bf \bar g}
_{\infty}$ has the form (3.18), although here of course $\bar g_{\bar \tau_{\infty}}$ might not be hyperbolic.

\medskip

 Thus, combining the discussion in Cases A and B, we see that all based limits of $(\bar \Sigma_{i}, g_{\bar \tau_{i}}, x_{i})$ are either complete hyperbolic manifolds, complete Seifert fibered spaces or complete $Sol$ manifolds, with a corresponding non-trivial group of isometries.

 To obtain the decomposition (0.8), any Riemannian 3-manifold $(\Sigma , g)$ has a thick-thin decomposition, i.e. for any fixed $\epsilon  > $ 0, we may write
\begin{equation} \label{e3.21}
\Sigma  = \Sigma^{\epsilon} \cup  \Sigma_{\epsilon}, 
\end{equation}
where $\Sigma^{\epsilon} = \{x\in\Sigma : vol_{g}B_{x}(1) \geq  \epsilon\}, \Sigma_{\epsilon} = \{x\in\Sigma : vol_{g}B_{x}(1) \leq  \epsilon\}.$ Of course these domains are not necessarily connected. Under a fixed curvature bound as in (3.13), this corresponds to a decomposition according to the size of the injectivity radius, as in Remark 1.6(ii). Now apply the decomposition (3.21) to each $(\bar \Sigma_{i}, \bar g_{\bar \tau_{i}}).$ By the discussion in Case A, for any fixed $\epsilon  > $ 0, the domains $(\bar \Sigma_{i}^{\epsilon}, \bar g_{\bar \tau_{i}}) \subset  (\bar \Sigma_{i}, \bar g_{\bar \tau_{i}}),$ when based at base point sequence $x_{i}\in\bar \Sigma_{i}^{\epsilon},$ (sub)-converge to domains $(\bar \Sigma_{\infty}^{\epsilon}, \bar g_{\infty})$. Now choose a sequence $\epsilon  = \epsilon_{j} \rightarrow $ 0, so that one now has a double sequence in $(i, j)$. By choosing a suitable diagonal subsequence $j = j(i)$, the domains $\bar \Sigma_{i}^{\epsilon_{j}}$ converge to the complete limit $(\bar \Sigma_{\infty}, \bar g_{\infty}) \subset  H$, and this defines the complete hyperbolic part $H$. On the other hand, for $\epsilon  = \epsilon_{j}$ sufficiently small, $\Sigma_{\epsilon_{j}}$ is a graph manifold, and this collapses everywhere as $\epsilon_{j} \rightarrow $ 0. This gives the decomposition (0.8).

\medskip

  We note that the transition from the $\epsilon_{j}$-thick part $\bar \Sigma_{i}^{\epsilon_{j}}$ to the $\epsilon_{j}$-thin part $(\bar \Sigma_{i})_{\epsilon_{j}}$ takes larger and larger diameter as $i$ and $j = j(i) \rightarrow  \infty ,$ so that the distance between these regions diverges to $\infty .$ Thus choosing different base points gives different limits only when the distance between base points goes to $\infty .$ Hence for instance if the diameter of $(\bar \Sigma_{i}, \bar g_{\tau_{i}})$ happens to remain uniformly bounded, then any limit has a unique geometry, i.e. independent of the base point.

 In fact, we see that the decomposition (3.21) is naturally refined into a further decomposition of $\Sigma_{\epsilon}$ according to the rank of the collapse, i.e. according to the type of group action of the limits described in Proposition 1.5; this is part of the general theory of collapse along F-structures, c.f. [14],[15]. Thus, based limits of $(\bar \Sigma_{i})_{\epsilon_{j}}$ which have a free isometric $S^{1}$ action become infinitely distant (in space) from based limits which have a free isometric $S^{1}\times S^{1}$ action. Hence one also obtains a decomposition of the graph manifold $G$ into Seifert fibered components, as discussed in \S 0.

 The decomposition (3.21) and hence (0.8) could change with different choices of sequences $\tau_{i} \rightarrow $ 0, as noted in \S 0; see also \S 3.3 for further discussion.

 This completes the proof of Theorem 3.1.

{\endproof}

\bbgin{remark} \label{r 3.2.}
  {\rm An alternate argument to the use of the volume comparison result (Corollary 1.3) in Case A can be given based on the monotonicity of the reduced Hamiltonian of Fischer-Moncrief [23]. Thus, it is proved in [23] that the function
$$\int_{\Sigma_{\tau}}H^{3}dV = \tau^{3}vol_{g_{\tau}}\Sigma_{\tau} $$
is monotone non-increasing in $\tau ,$ and is constant if and only if $\Sigma $ is hyperbolic.}
\end{remark}

\bbgin{remark} \label{r 3.3.}
  {\rm Suppose (3.20) is strengthened to the statement that $|H(\tau_{i})|\cdot  t_{min}(\tau_{i}) \rightarrow $ 3. Then together with (1.15) and fact that $H$ is constant, it follows that $t_{max}/t_{min} \rightarrow $ 1 on $\Sigma_{i} = \Sigma_{\tau_{i}}.$ This means that the corresponding rescaled proper time $\bar t_{i}$ following (3.10) converges to 1 everywhere on $\bar \Sigma_{i}$ which in turn implies that the lapse $\bar \alpha_{i}$ on $\bar \Sigma_{i}$ also approaches a constant function everywhere on $\bar \Sigma_{i}.$

 It follows that the limit $\bar \Sigma_{\infty}$ of $\bar \Sigma_{i}$ is hyperbolic everywhere so that $(\bar \Sigma_{\infty}, \bar g_{\infty})$ is a compact hyperbolic manifold, diffeomorphic to $\Sigma .$

 We claim that in this situation, the rescalings $(\bar \Sigma_{\bar \tau}, \bar g_{\bar \tau})$ converge to the hyperbolic limit $(\Sigma , \bar g_{\infty}),$ for {\it all} sequences $\tau  \rightarrow $ 0, i.e. by Mostow rigidity, the limit is unique. For the monotonicity of (3.17) implies that any limit of a sequence gives rise to a complete hyperbolic manifold $H' $ embedded in $\Sigma ,$ as in (0.8). Now Thurston's cusp closing theorem, c.f. [38], implies that the volume of any hyperbolic cusp $H' $ embedded in $\Sigma $ has volume strictly larger than that of the compact hyperbolic metric $(\Sigma , \bar g_{\infty}).$ Since the volume of the graph manifold part $G$ in (0.8) has non-negative volume in the limit, the claim follows from the volume monotonicity.}
\end{remark}

{\bf \S 3.2.}
 In this section, we prove that the assumption (3.4) always holds, at least when the curvature assumption (3.3) is strengthened to (0.11).

\bbgin{theorem} \label{t 3.4.}
  Let ({\bf M, g}) be a cosmological CMC space-time satisfying the curvature assumption (0.11) and {\bf M}$ = M_{{\cal F}}.$ Then there exists a constant $\delta  = \delta ({\bf M, g}) >  0$ such that
\begin{equation} \label{e3.22}
|H| \geq  \delta /t_{min}. 
\end{equation}
\end{theorem}

{\bf Proof:}
 The proof will proceed in several steps, but overall the proof proceeds by contradiction, and so we suppose there exists a sequence $\tau_{i} \rightarrow $ 0, and so $t_{min}(\tau_{i}) \rightarrow  \infty ,$ such that
\begin{equation} \label{e3.23}
t_{min}(\tau_{i})\cdot |H_{\Sigma_{\tau_{i}}}| \rightarrow  0, \ \ {\rm as} \ \  i \rightarrow  \infty . 
\end{equation}

 Let $x_{i}\in\Sigma_{i}$ be base points realizing $t_{min},$ so $t(x_{i}) = t_{min}(\tau_{i}).$ In contrast to (3.8), throughout this section we consider the rescaled space-time metrics
\begin{equation} \label{e3.24}
{\bf \bar g}
_{i} = t_{min}(\tau_{i})^{-2}\cdot {\bf g}
, 
\end{equation}
with renormalized proper time $\bar t_{i} = t/t_{min}(\tau_{i})$ and CMC time $\bar \tau_{i} = \tau\cdot  t_{min}(\tau_{i}).$ As in the proof of Theorem 3.1, the argument in (3.12) shows that the curvature ${\bf \bar R}
_{i}$ of ${\bf \bar g}
_{i}$ is uniformly bounded in the region where $\bar t_{i} \geq  t_{o},$ for any given $t_{o} > $ 0. It then follows from the arguments in \S3.1, c.f. the discussion preceding (3.18), that the space-times ({\bf M}, ${\bf \bar g}_{i}, x_{i})$ have a subsequence converging in the weak $L^{2,p}$ topology to a limit $L^{2,p}\cap C^{1,\beta}$ maximal vacuum space-time $({\bf M}
_{\infty}, {\bf \bar g}
_{\infty}, x_{\infty}),$ where one must pass to covers in the case of collapse. In this latter case, $({\bf M}
_{\infty}, {\bf \bar g}
_{\infty})$ has at least a free isometric space-like $S^{1}$ action. The parameters $\bar t_{i}$ and $\bar \tau_{i}$ converge to limit parameters $\bar t_{\infty}$ and $\bar \tau_{\infty}.$

 The CMC surfaces $\Sigma_{\tau_{i}},$ now labeled as $\bar \Sigma_{i},$ are such that a subsequence of $(\bar \Sigma_{i}, \bar g_{\tau_{i}}, x_{i})$ also converges in the weak $L^{2,p}$ topology, and uniformly on compact subsets, to a limit $L^{2,p}\cap C^{1,\beta}$ CMC hypersurface $(\bar \Sigma_{\infty}, \bar g_{\infty}, x_{\infty}).$ Of course, by construction, $\bar t_{\infty} \geq $ 1 on $\bar \Sigma_{\infty}.$ Similarly, the CMC foliation $M_{\bar {\cal F}_{i}}$ in the scale (3.24) converges to a limit $L^{2,p}\cap C^{1,\beta}$ CMC foliation $M_{\bar {\cal F}_{\infty}}$ of $({\bf M}_{\infty}, {\bf \bar g}_{\infty})$ with leaves $\bar \Sigma_{\bar \tau_{\infty}}.$ Again, in the case of collapse, the leaves are unwrapped and so have at least an isometric $S^{1}$ action.

 The assumption on $|\nabla{\bf R}|$ in (0.11) implies in the same way that $|\nabla{\bf R_{\bar g_{i}}}|$ is uniformly bounded where $\bar t_{i} \geq  t_{o}.$ By elliptic regularity applied to the equation (2.2), it follows that the intrinsic curvature $Ric_{\bar g_{i}}$ and extrinsic curvature $K_{\bar g_{i}}$ of the leaves are uniformly bounded in $L^{3,p}$ and $L^{2,p}$ respectively, (compare with the proof of Theorem 0.1). Thus, the convergence above is everywhere in the weak $L^{3,p}$ topology and the limits are in $L^{3,p}\cap C^{2,\beta}.$

 Now the scale-invariance of (3.23) and the smoothness of the convergence imply that the limit $\bar \Sigma_{\infty}$ is a complete {\it maximal} hypersurface, i.e. $H =$ 0. In fact, since $\tau  = H$ is monotone decreasing on ({\bf M, g}), it follows that all the leaves $\bar \Sigma_{\bar \tau_{\infty}}$ of $M_{\bar {\cal F}_{i}}$ are complete maximal hypersurfaces. In particular, the limit mean curvature parameter $\bar \tau_{\infty}$ is identically 0.

 This means that the lapse function $\alpha $ on ({\bf M, g}) satisfies
\begin{equation} \label{e3.25}
\alpha (x_{i}) >>  t_{min}(\tau_{i})^{2}, 
\end{equation}
so that the corresponding renormalized lapse functions $\bar \alpha_{i} = \alpha /\alpha (x_{i})$ on ({\bf M}, ${\bf \bar g}_{i}, x_{i})$ satisfy $\bar \alpha_{i}(x_{i}) >> $ 1.  To remedy this situation, we redefine the normalized lapse by setting
\begin{equation} \label{e3.26}
\bar \alpha_{i} = \frac{\alpha}{\alpha (x_{i})}. 
\end{equation}
The estimates (2.21) imply that $\bar \alpha_{i}$ is uniformly bounded within $\bar g_{i}$ bounded distance to $x_{i}$ on $\bar \Sigma_{i}.$ Similarly, the arguments preceding (3.18) imply that $\bar \alpha_{i}$ is uniformly bounded on bounded domains on leaves within bounded proper time distance to $x_{i}.$ Hence $\bar \alpha_{i}$ converges to the limit lapse function $\bar \alpha_{\infty}$ on $({\bf \bar M}_{\infty}, {\bf \bar g}_{\infty}).$ Similarly, we redefine the CMC time parameter by
\begin{equation} \label{e3.27}
\bar \tau_{i} = (\alpha (x_{i}))^{1/2}\cdot \tau . 
\end{equation}
Then $\bar \tau_{i}$ converges smoothly to a parametrization $\bar \tau_{\infty}$ of the leaves of the limit foliation. It follows that the limit lapse function $\bar \alpha_{\infty}$ satisfies the following lapse equation
\begin{equation} \label{e3.28}
-\Delta \alpha  + |K|^{2}\alpha  = 0, 
\end{equation}
on each leaf $\bar \Sigma_{\bar \tau_{\infty}},$ where $\alpha  = \bar \alpha_{\infty}.$ The equation (3.28) means that the variation vector field $\alpha T,$ where $T$ is the unit normal, is a Jacobi field for the volume functional, i.e. the flow of $\alpha T$ preserves the volume of (domains on) the leaves.

 With respect to these redefinitions of $\bar \alpha_{\infty}$ and $\bar \tau_{\infty},$ the limit space-time $({\bf \bar M}
_{\infty}, {\bf \bar g}
_{\infty})$ has the form (3.18).

\medskip

 Our task is now to rule out this limiting or near limiting behavior of the space-time ({\bf M, g}). The key to this is the following result.

\bbgin{lemma} \label{l 3.5.}
  The limit maximal hypersurface $\bar \Sigma_{\infty}$ cannot be flat, i.e.
\begin{equation} \label{e3.29}
|Ric|_{\bar g_{\infty}}(y) >  0 
\end{equation}
for some $y\in\bar \Sigma_{\infty}.$
\end{lemma}

{\bf Proof:}
 Assuming $(\Sigma_{\infty}, g_{\infty})$ is flat, we will obtain a contradiction by the Cauchy stability theorem. Thus, suppose $(\Sigma_{\infty}, g_{\infty})$ is flat. By the constraint equation (1.2), since $\bar s_{\infty} =$ 0 and $H =$ 0, we have
$$\bar K_{\infty} = 0, $$
so that $\bar \Sigma_{\infty}$ is totally geodesic, (i.e. time-symmetric).

 It follows that for $i$ sufficiently large, the CMC surfaces $(\bar \Sigma_{i}, \bar g_{i}, x_{i})$ are almost flat and totally geodesic, i.e.
\begin{equation} \label{e3.30}
|Ric|_{\bar g_{i}}(y) <  \epsilon ,  \ \ {\rm and} \ \  |K|_{{\bf \bar g_{i}}
}(y) <  \epsilon , 
\end{equation}
for all $y\in B_{x_{i}}(R),$ where $R$ may be made arbitrarily large and $\epsilon  > $ 0 arbitrarily small by choosing $i$ sufficiently large; here $B_{x_{i}}(R)$ is the geodesic $R$-ball in $(\bar \Sigma_{i}, \bar g_{i})$ about $x_{i}.$ In fact, as discussed above following (3.24), by Remark 1.6(i), the curvature bounds (0.11) imply that $(B_{x_{i}}(R), \bar g_{i})$ is $\epsilon$-close in the weak $L^{3,p}$ topology to the flat metric, and $K$ is $L^{2,p}$ close to the 0-form.

 It follows that the Cauchy data $(\bar g_{i}, K_{i})$ on the CMC surfaces $\bar \Sigma_{i}$ in the region $(B_{x_{i}}(R), \bar g_{i})$ are $\epsilon$-close in weak $(L^{3,p}, L^{2,p})$ to trivial Cauchy data. By Sobolev embedding, $L^{3,p}$ is compactly contained in $H^{s},$ for any $s < $ 3, where $H^{s}$ is the Sobolev space with $s$ derivatives in $L^{2}.$ Similarly, $L^{2,p}$ is compactly contained in $H^{s-1}.$ This means that the Cauchy data $(\bar g_{i}, K_{i})$ are strongly close to trivial data in $(H^{s}, H^{s-1}), s < $ 3.

 Choosing $s > $ 2.5, the Cauchy stability theorem, c.f. [23], then implies that the maximal Cauchy development of $B_{x_{i}}(R/2) \subset  \bar \Sigma_{i}$ to the past exists for a proper time $T = T(\epsilon ),$ where $T$ may be made arbitrarily large if $\epsilon $ is chosen sufficiently small. However, $\bar t_{i}(x_{i}) =$ 1 and by the remarks in \S 0, the space-time ({\bf M, g}) has a singularity, i.e. fails to be globally hyperbolic, within {\bf g}-bounded proper time to the past of $\Sigma_{\tau_{o}}.$ Hence $({\bf M, \bar g_{i}})$ has a singularity to the past of $\bar \Sigma_{i}$ within proper time at most 2, for $i$ large. This contradiction proves the result.

{\endproof}

\bbgin{remark} \label{r 3.6.}
  {\rm The following generalization of Lemma 3.5 will be needed in the work to follow. Namely, the proof above leads to the same contradiction if there is an $\epsilon  > $ 0 sufficiently small, a number $D = D(\epsilon )$ sufficiently large, and a point $y\in\bar \Sigma_{\infty}$ such that the metric
\begin{equation} \label{e3.31}
g'  = \bar t_{\infty}(y)^{-2}\cdot \bar g_{\infty} 
\end{equation}
is $\epsilon$-close to the flat metric in the $H^{s}$ topology, $s > $ 2.5, in the ball $(B_{y}(D), g' ).$ 

 We remark that this use of the Cauchy stability theorem is the only place in the proof that the assumption on the derivative of the curvature in (0.11) is needed.}
\end{remark}

 Thus, to prove Theorem 3.4, it suffices to prove the limit maximal surface $(\bar \Sigma_{\infty}, \bar g_{\infty})$ must either be flat or have a point $y$ satisfying the weaker assumption in Remark 3.6. The proof of this needs to be divided into non-collapse and collapse cases, as in \S 3.1.

\medskip

{\bf Case A. (Non-Collapse).}

 Since the space-time $({\bf \bar M}_{\infty}, {\bf \bar g}_{\infty})$ is a rescaled limit at infinity of $({\bf M, g})$, there exists a future directed time-like geodesic ray $\gamma $ in $({\bf M, g})$ whose rescalings in the metrics (3.24) converge to a geodesic ray $\bar \gamma_{\infty}$ in $({\bf \bar M}_{\infty}, {\bf \bar g}_{\infty}).$ As in the proof of Proposition 1.2, let $S(s) = t^{-1}(s)$ be the time $s$ geodesic sphere about $\Sigma_{\tau_{o}}$ in ({\bf M, g}) and let $\bar S_{i}(s) = \bar t_{i}^{-1}(s)$ be its rescaling in $({\bf \bar M}_{i}, {\bf \bar g}_{i})$ from (3.24). Let $z_{i} = \gamma (t_{i})\in S(t_{i})$, where $t_{i} = t(x_{i})$. Thus $z_{i} \in \bar S_{i}(1)$ and $x_{i}\in\bar S_{i}(1).$

 The geodesic spheres $(\bar S_{i}(1), z_{i})$ converge, (in subsequences), to a Lipschitz, achronal surface $\bar S_{\infty}(1) = \bar t_{\infty}^{-1}(1) \subset  ({\bf \bar M}_{\infty}, {\bf \bar g}_{\infty})$ containing the base point $x_{\infty}.$ The limit proper time function $\bar t_{\infty}$ on $({\bf \bar M}_{\infty}, {\bf \bar g}_{\infty})$ induces a Lipschitz 3+1 splitting of the limit space-time metric ${\bf \bar g}_{\infty},$ (which of course is not well-defined on the cut locus of $\bar t_{\infty}).$

 For any fixed $R <  \infty ,$ let $D_{z_{i}}(R\cdot  t_{i})$ be the geodesic ball of radius $R\cdot  t_{i}$ about $z_{i}$ in $S(t_{i}),$ and $\bar D_{z_{i}}(R)$ be its rescaling in $\bar S_{i}(1)$.

 In this case, we assume that there exists $\nu_{o} > $ 0 such that
\begin{equation} \label{e3.32}
vol \bar D_{z_{i}}(1) \geq  \nu_{o} >  0, \ \ {\rm as} \ \ i \rightarrow  \infty , 
\end{equation}
so that the domains $\bar D_{z_{i}}(1)$ are non-collapsing and hence converge, (in the Hausdorff topology), to the limit domain $\bar D_{z_{\infty}}(1) \subset  \bar S_{\infty}(1).$ 

  Now as in (1.16), the expansion $\theta $ of the congruence formed by time-like geodesics normal to the spheres $S(s) \subset ({\bf M, g})$ satisfies (1.15), (see the proof of Proposition 1.2). Hence, if $dV_{\sigma}(s)$ denotes the infinitesimal volume of the family $S(s)$ along any geodesic $\sigma$ in this congruence, then $dV_{\sigma}(s)/s^{3}$ is monotone non-increasing as $s \rightarrow  \infty .$

 It then follows from (3.32) as with (3.19) that the domain in $({\bf \bar M}_{\infty}, {\bf \bar g}_{\infty})$ formed by the geodesics normal to $\bar D_{z_{\infty}}(1) \subset  \bar S_{\infty}(1)$ is a volume cone, and hence this domain is contained in a flat Lorentz cone as in (0.12). This flat structure extends past the cone on $\bar D_{z_{i}}(1)$ and implies that all of $({\bf \bar M}_{\infty}, {\bf \bar g}_{\infty})$ is a flat Lorentz cone. In particular, the universal cover of $({\bf \bar M}_{\infty}, {\bf \bar g}_{\infty})$ is flat Minkowski space $({\Bbb R}^{4}, \eta ).$

 Now the limit surface $(\Sigma_{\infty}, g_{\infty})$ is a complete maximal hypersurface in $({\bf \bar M}
_{\infty}, {\bf \bar g}
_{\infty}),$ which lifts to a complete maximal hypersurface in $({\Bbb R}^{4}, \eta ).$ It is an easy consequence of the maximum principle applied to (2.4), as in the proof of Proposition 2.2, that the only complete maximal hypersurfaces in $({\Bbb R}^{4}, \eta )$ are flat and totally geodesic; this is also proved in [16]. Hence $\bar \Sigma_{\infty} \subset  ({\bf \bar M}
_{\infty}, {\bf \bar g}
_{\infty})$ is flat. Lemma 3.5 thus rules out the possibility of this case.

\medskip

{\bf Case B. (Collapse).}

 In this case, we suppose (3.32) does not hold, so that
\begin{equation} \label{e3.33}
vol \bar D_{z_{i}}(1) \rightarrow  0, \ \ {\rm as} \ \ i \rightarrow  \infty . 
\end{equation}
It follows from the monotonicity used above that for any fixed $s_{o} > $ 0, and $s\in [s_{o}, s_{o}^{-1}],$ the domains $\bar D_{z_{i}}(s) \subset  \bar S_{i}(s)$ also satisfy (3.33).

 This implies that the CMC leaf $\bar \Sigma_{i}' $ containing $z_{i}$ is also collapsing when based at $z_{i}$ as $i \rightarrow \infty$. By the uniform control on the geometry following (3.24), it follows that all CMC leaves within ${\bf \bar g}_{i}$ bounded distance, in space and proper time, to $(\bar \Sigma_{i}' , z_{i})$, are collapsing as $i \rightarrow  \infty .$ In particular, the leaves $\bar \Sigma_{i}$ containing the base points $x_{i}$ are collapsing everywhere within bounded distance to $x_{i}.$

 Thus, as described before, we unwrap the collapse of $\bar \Sigma_{i}$ and the space-times to obtain a limit $\bar \Sigma_{\infty}$ and limit space-time $({\bf \bar M}_{\infty}, {\bf \bar g}_{\infty}, x_{\infty}).$ These limits have a non-trivial group of isometries, and ${\bf \bar M}_{\infty}$ has a foliation by maximal hypersurfaces.

 Now unfortunately, we need to divide the discussion into two further subcases, according to the size of the isometry group.

{\bf Case B(I). $(S^{1}\times S^{1})$.}

 Suppose the limit $(\bar \Sigma_{\infty}, \bar g_{\infty})$ has a free isometric $S^{1}\times S^{1}$ action. By the constraint equation (1.2), the metric $\bar g_{\infty}$ is a complete metric of non-negative scalar curvature on $\bar \Sigma_{\infty}.$ Further, any orbit of the $S^{1}\times S^{1}$ action on $\bar \Sigma_{\infty}$ is an incompressible torus in $\bar \Sigma_{\infty}.$ However, by a result of [26], any complete 3-manifold of non-negative scalar curvature which has an incompressible torus must be flat. Hence we see that $(\bar \Sigma_{\infty}, \bar g_{\infty})$ is in fact flat. Lemma 3.5 again rules out this possibility. Of course the same arguments apply, (and are even easier), if the limit $\bar \Sigma_{\infty}$ has a locally free $T^{3}$ or $Nil$ action.

{\bf Case B(II). $(S^{1})$.}

 Suppose the limit $(\bar \Sigma_{\infty}, \bar g_{\infty})$ has (at most) a free isometric $S^{1}$ action. Thus, we may assume that $\bar \Sigma_{\infty}$ is an $S^{1}$ bundle over a surface $V$, with induced complete Riemannian metric $g_{V}.$ By the result of [26] above, it follows that $V$ must be simply connected, and hence topologically $\bar \Sigma_{\infty}$ is a solid torus $D^{2}\times S^{1}.$

 In this case, we are not able to prove directly that $(\bar \Sigma_{\infty}, \bar g_{\infty})$ is flat, since a solid torus carries many complete metrics of positive scalar curvature. Instead, we will prove that the hypothesis concerning (3.31) in Remark 3.6 holds, which again gives a contradiction.

 To do this, we need to understand the asymptotic geometry of $(\bar \Sigma_{\infty}, \bar g_{\infty}).$ To simplify, we drop the bar and subscript everywhere from the notation; thus let $(\Sigma , g)$ denote $(\bar \Sigma_{\infty}, \bar g_{\infty}),$ $({\bf M, g})$ denote $({\bf \bar M_{\infty}, \bar g_{\infty}}),$ $t(x)$ denote $\bar t_{\infty}(x),$ etc. The scale-invariant bound (0.11) gives
\begin{equation} \label{e3.34}
|{\bf R}
| + t|\nabla{\bf R}
| \leq  \frac{C}{t^{2}}, 
\end{equation}
on $(\Sigma , g)$. Since by construction $t \geq $ 1 on $\Sigma ,$ the curvature {\bf R}
 and its derivative $\nabla{\bf R}
$ are thus uniformly bounded on $\Sigma .$ By Proposition 2.2 it follows that
\begin{equation} \label{e3.35}
|Ric| \leq  \frac{C}{t^{2}}, |K| \leq  \frac{C}{t}, 
\end{equation}
on $(\Sigma , g)$. One derives (3.35) most easily from (3.34) by working in the scale where $t =$ 1 and using scale-invariance. In the same way, applying elliptic regularity to the equation (2.2) gives the scale-invariant bounds
\begin{equation} \label{e3.36}
t^{-3/p}|t^{3}\nabla Ric|_{L^{p}} \leq  C, t^{-3/p}|t^{3}\nabla^{2}K|_{L^{p}} \leq  C, 
\end{equation}
where the $L^{p}$ norms are taken over regions of diameter $\frac{1}{2}t$ in $(\Sigma , g)$. In the case of collapse, the $L^{p}$ norms are understood to be taken in the corresponding covering spaces.

 Let $r: \Sigma  \rightarrow  {\Bbb R} $ be the (Riemannian) distance function on $(\Sigma , g)$ from the given base point $x_{\infty}\in\Sigma .$ We need the following Lemma.

\bbgin{lemma} \label{l 3.7.}
  The functions $t$ and $r$ on $(\Sigma ,$ g) satisfy
\begin{equation} \label{e3.37}
t <<  r, 
\end{equation}
on any sequence $y_{i}$ with $r(y_{i}) \rightarrow  \infty .$
\end{lemma}

{\bf Proof:}
 If $t(y_{i})$ is bounded, then (3.37) is obvious, so suppose $t(y_{i}) \rightarrow  \infty $ as $r(y_{i}) \rightarrow  \infty .$ Then by (3.34)-(3.35) the metric $g$ on $\Sigma $ becomes flat everywhere as $t \rightarrow  \infty $ and ({\bf M, g}) thus approaches empty Minkowski space $({\Bbb R}^{4}, \eta ),$ (or a discrete quotient of it). The maximal foliation of ({\bf M, g}) approaches the constant foliation of $({\Bbb R}^{4}, \eta )$ where all leaves are parallel, i.e. time-equidistant, since this is the only maximal foliation of $({\Bbb R}^{4}, \eta)$, c.f. [16].

 Hence, for $i$ sufficiently large, the lapse function $\alpha  = \bar \alpha_{\infty},$ (c.f. (3.28)), approaches a constant function on domains of fixed diameter about the base point $y_{i}$ while the light cones of $({\bf M, g})$ have axis becoming perpendicular to the leaves. This means that the time function $t$ also approaches a constant on such domains. Since $r$ obviously continues to increase linearly, (3.37) follows.

{\endproof}

 We are now ready to examine the asymptotic structure of $(\Sigma , g)$. Thus for any sequence $y_{i}$ in $\Sigma $ with $r(y_{i}) \rightarrow  \infty ,$ consider the rescaled metrics
\begin{equation} \label{e3.38}
g_{i} = t(y_{i})^{-2}\cdot  g. 
\end{equation}
Hence, $t_{i}(y) = t(y)/t(y_{i})$ satisfies $t_{i}(y_{i}) =$ 1 and by Lemma 3.7, $r_{i}(y) = r(y)/r(y_{i}) \rightarrow  \infty .$ Thus the Riemannian distance of $y_{i}$ to the base point $x_{\infty}$ of $\Sigma$ in the $g_{i}$ metric diverges to $\infty $, while the time function $t_{i}$ remains uniformly bounded at $y_{i}$. We may alter the choice of $y_{i}$ if necessary so that it satisfies all of the estimates above and in addition
\begin{equation} \label{e3.39}
t_{i}(z_{i}) \geq  \frac{1}{2}, 
\end{equation}
for all $z_{i}\in\Sigma $ such that $dist_{g_{i}}(y_{i}, z_{i}) \leq  C$, where $C$ may be made arbitrarily large if $i$ is sufficiently large. To see this, for any sequence $y_{i}$ with $r(y_{i}) \rightarrow \infty$, let $\rho_{i} = r(y_{i})$ and let $A_{i} = A(\frac{1}{2}\rho_{i}, 2\rho_{i})$ be the $g$-geodesic annulus centered at $x_{\infty}$ of inner and outer radii $\frac{1}{2}\rho_{i}, 2\rho_{i}.$ Choose points $q_{i}\in A_{i}$ realizing the minimal value of the scale-invariant ratio $t(q)/dist_{g}(q, \partial A_{i}).$ By (3.37), this minimal value converges to 0 as $i \rightarrow  \infty $ and relabeling $q_{i}$ to $y_{i}$ easily gives the estimate (3.39).

 Thus, by (3.36), the metrics $g_{i}$ based at $y_{i}$ have a subsequence converging in the weak $L^{3,p}$ topology, uniformly on compact subsets to a complete $L^{3,p}$ limit $(\Sigma^{\infty}, g^{\infty}),$ unwrapping in the case of collapse. We claim that any such limit must be flat. 

  The proof of this is essentially the same as in [5, Prop. 4.1-4.3], to which we refer for some further details. Suppose first the metrics $(g_{i})_{V}$ on the {\it base} space $V$ of $(\Sigma , g_{i}, y_{i})$ are collapsing at $y_{i}.$ By unwrapping this collapse, (and any possible collapse in the invariant $S^{1}$ fiber direction), the limit $(\Sigma^{\infty}, g^{\infty})$ is thus a complete metric of non-negative scalar curvature on the limit manifold $\Sigma^{\infty} = T^{2}\times {\Bbb R} .$ As above, by [26], the metric $g^{\infty}$ must be flat; this is the same argument as in [5, Prop.4.3, Case II]. 

  If the base metrics $(g_{i})_{V}$ are not collapsing, then the limit $(\Sigma^{\infty}, g^{\infty})$ is complete and of non-negative scalar curvature on ${\Bbb R}^{2}\times S^{1}.$ In this case, the proof in [5, Prop.4.3, Case I], together with the control given by (3.36), proves again that $(\Sigma^{\infty}, g^{\infty})$ is flat. Here we note that although [5, Prop.4.3] applies to complete $S^{1}$ invariant metrics on ${\Bbb R}^{2}\times S^{1}$ of non-negative scalar curvature which satisfy certain elliptic equations, (the ${\cal Z}_{c}^{2}$ equations), the proof only requires these equations to obtain regularity estimates; in the case at hand, these are provided by (3.36). This proves the claim.

 By Sobolev embedding, the convergence to the limit $(\Sigma^{\infty}, g^{\infty})$ is in the strong $H^{s}$ topology, for any $s < $ 3. Hence, since the limit is flat, there are arbitrarily large domains in $(\Sigma , g_{i}, y_{i})$ on which $g_{i}$ is $\epsilon$-close to a flat metric in the $H^{s}$ topology.

 Thus, the hypothesis of Lemma 3.6 is satisfied, and one obtains a contradiction from the Cauchy stability theorem as in Lemma 3.5.

 This completes the proof of Case B(II), which thus also completes the proof of Theorem 3.4. Together with Theorem 3.1, this also completes the proof of Theorem 0.3.

{\endproof}

{\bf \S 3.3. Discussion on collapse behavior.}

{\bf (I).}
 The collapse behavior arising in Theorems 0.3 and 3.1, c.f. \S 3.1 Case B, does actually occur in numerous examples. In fact, let ({\bf M, g}) be a Bianchi space-time, i.e. ({\bf M, g}) locally has a 3-dimensional local isometry group whose action is free and space-like. All such space-times admit a global CMC foliation ${\cal F} $ whose leaves $\Sigma_{\tau}$ are invariant under the action, i.e. each leaf $\Sigma_{\tau}$ is locally homogeneous. Further, the foliation ${\cal F} $ exists for all allowable CMC time, for any value of $\sigma (\Sigma ),$ and $M = M_{{\cal F}},$ c.f. [19].

 The Bianchi space-times correspond very closely to the eight Thurston geometries arising in the geometrization program. In particular, all eight of the Thurston geometries occur as space-like Cauchy surfaces in a corresponding Bianchi space-time, with the interesting exception of $S^{2}\times {\Bbb R} ,$ which corresponds to a Kantowski-Sacks space-time. We refer to [6] for a discussion of this Bianchi-Thurston correspondence and for further references. 

 The five geometries ${\Bbb R}^{3},$ $Nil$, $H^{2}\times {\Bbb R} ,$ SL(2, ${\Bbb R} )$ and $Sol$ all give 3-manifolds $\Sigma $ with $\sigma (\Sigma ) =$ 0 and the corresponding Bianchi space-times are future geodesically complete. The curvature assumption (0.11) is satisfied and the space-times collapse in the sense of \S 3.1, Case B, as $\tau  \rightarrow $ 0 or $t_{\tau} \rightarrow  \infty .$

 The simplest example is the Kasner metric, given by
\begin{equation} \label{e3.40}
{\bf g} = - dt^{2} + \sum_{i=1}^{3}t^{2p_{i}}d\theta_{i}\otimes d\theta_{i}, 
\end{equation}
where $\{p_{i}\}$ are any constants satisfying $\sum p_{i} =$ 1, $\sum p_{i}^{2} =$ 1 and $\theta_{i}$ are parameters for $T^{3}.$ This has ${\Bbb R}^{3}$ geometry, with $\Sigma $ a flat 3-torus, $H = \tau  = - t^{-1}.$ In this case, for any sequence $\tau_{i} \rightarrow $ 0, the rescalings $\bar \Sigma_{i}$ of $\Sigma_{i}$ as in (3.9) collapse with bounded curvature and diameter converging to 0. Hence, the unwrapping in covers give rise to a $T^{3}$ action on the limit surface, i.e. one obtains the Kasner metric again, (with the same exponents).

 For the remaining 3 geometries, the $S^{3}$ and $S^{2}\times {\Bbb R} $ geometries have $\sigma (\Sigma ) > $ 0 and give Bianchi and Kantowski-Sachs space-times which recollapse, as in the recollapse conjecture, while the hyperbolic geometry $H^{3}(-1)$ gives the flat Lorentz cone (0.12) and so evolves just by rescalings.

{\bf (II).}
 As discussed in \S 3.1 Case B, if the rescalings (3.9) collapse at $x_{i},$ then one may unwrap the collapse of both the space and space-time, and pass to a limit space-time $({\bf \bar M}
_{\infty}, {\bf \bar g}
_{\infty}, x_{\infty}),$ with geodesically complete CMC Cauchy surface $(\bar \Sigma_{\infty}, \bar g_{\infty}, x_{\infty})$ and a corresponding limit CMC foliation $\bar M_{\bar {\cal F}_{\infty}}.$ The leaves $\bar \Sigma_{\bar \tau_{\infty}}$ of $\bar M_{\bar {\cal F}_{\infty}}$ have at least a free isometric $S^{1}$ action, and possibly a locally free isometric action of a 2 or 3-dimensional group. The leaves of $\bar M_{\bar {\cal F}_{\infty}}$ may be either compact, or complete and non-compact. The limit space-time $({\bf \bar M}_{\infty}, {\bf \bar g}_{\infty})$ is time-like geodesically complete to the future of $\bar \Sigma_{\infty}.$

 In the case of an $S^{1}\times S^{1}$ action, these limit space-times $({\bf \bar M}
_{\infty}, {\bf \bar g}
_{\infty})$ include the Gowdy space-times and have been quite well studied, at least when $\bar \Sigma_{\infty}$ is compact, c.f. [6] and references therein.

 One can now study (again) the further long-time evolution of these limit space-times. It is an interesting open question whether these limits $({\bf \bar M}
_{\infty}, {\bf \bar g}
_{\infty})$ evolve as $\bar t_{\infty} \rightarrow  \infty $ to a Bianchi space-time as in (I), and if so whether in fact the space-times $({\bf \bar M}
_{\infty}, {\bf \bar g}
_{\infty})$ themselves are already Bianchi. To prove this, it seems likely one would need some kind of monotonicity structure to replace the volume monotonicity of Corollary 1.3, (or the Hamiltonian monotonicity [23]), used in the proof of Theorem 0.3.

{\bf (III).}
 Finally, we discuss the relation between weak and strong geometrizations, and the recollapse conjecture. 

 Let $\Sigma  = H \cup  G$ be a weak geometric decomposition as given by Theorem 0.3 or Theorem 3.1. The manifold $G$ has toral boundary ${\cal T}  = \cup T_{i}$ and the decomposition (0.8) is strong, and consequently unique, if each torus $T_{i}$ is incompressible in $\Sigma .$ Now of course since the hyperbolic part $H$ has no incompressible tori, $T_{i}$ is incompressible in $\Sigma $ if and only if $T_{i}$ is incompressible in $G$.

 It is essentially a standard result in 3-manifold topology, c.f. [30, Ch.II.2], [35], that each boundary component $T_{i}$ is incompressible in $G$ if and only if $G$ has no solid tori factors, i.e. in the decomposition of $G$ into Seifert fibered spaces $S_{i}$ discussed in \S 0, no $S_{i}$ is a solid torus $D^{2}\times S^{1}.$ 

 Thus, the question is whether solid torus factors $D^{2}\times S^{1}$ can arise asymptotically from the collapse behavior, (compare with Case B(II) in the proof of Theorem 3.4). At least in a special case, this question is closely related to the recollapse conjecture. 

 Thus suppose one has a $D^{2}\times S^{1}$ factor in $G$ arising from the asymptotic collapsing geometry of ({\bf M, g}) as discussed in the proof of Theorem 3.1. The limiting metric $\bar g_{\infty}$ on $D^{2}\times S^{1}$ is complete and invariant under a free isometric $S^{1}$ action.

 Suppose, for some $t_{\tau_{i}}$ sufficiently large, i.e. $\tau_{i}$ sufficiently small, there is $D^{2}\times S^{1}$ region contained in $(\bar \Sigma_{i}, \bar g_{\tau_{i}})$ with a totally geodesic boundary $T^{2}$ in the $\bar g_{i}$ metric. One can then double or reflect across the boundary to obtain a closed 3-manifold $S^{2}\times S^{1}$ with an isometric ${\Bbb Z}_{2}$ action, fixing the boundary. Of course in this process we have created a different space-time. Suppose further the extrinsic curvature $\bar K$ of $\bar \Sigma_{i}$ at the boundary $T^{2}$ is invariant under this ${\Bbb Z}_{2}$ action so that the Cauchy data on $\bar \Sigma_{i}$ are ${\Bbb Z}_{2}$ invariant. Now $\sigma (S^{2}\times S^{1}) > $ 0 and so the recollapse conjecture, if true, implies that this $S^{2}\times S^{1}$ must evolve into a singularity in finite future time. Since the evolution of Cauchy data preserves isometries, this implies that the $D^{2}\times S^{1}$ factor in $(\bar \Sigma_{i}, \bar g_{\tau_{i}})$ also evolves to a singularity in finite future time. This is of course a contradiction.

 It would of course be interesting if this argument could be strengthened to prove in general that the recollapse conjecture implies that a weak decomposition of $\Sigma $ must necessarily be a strong decomposition.

\section{The Future Boundary of $M_{{\cal F}}$ in {\bf M}.}
\setcounter{equation}{0}

 In this section, we discuss the situation where $M_{{\cal F}}$ is strictly contained in {\bf M}, c.f. \S 0. In this case, Theorem 0.1 implies that {\bf M}
 has global existence in CMC time both to the past $\tau  \rightarrow  -\infty $ and to the future $\tau  \rightarrow $ 0, (or $\tau  \rightarrow  +\infty $ when $\sigma (\Sigma ) > $ 0). In particular, by the discussion at the end of \S 2, the past singularity is a crushing singularity. Thus, in this section, we only consider the boundary
\begin{equation} \label{e4.1}
\partial_{o}M_{{\cal F}} \subset  {\bf M}
\end{equation}
of $M_{{\cal F}}$ formed when $\tau  \rightarrow $ 0, i.e. the boundary to the future of the initial surface $\Sigma_{\tau_{o}}$ from (0.2). The condition (4.1) means that the space-time ({\bf M, g}) i.e. the maximal Cauchy development of initial data on $\Sigma ,$ extends a definite amount locally, i.e. in a neighborhood of any point in $\partial_{o}M_{{\cal F}},$ to the future of the CMC foliation ${\cal F} ,$ so that $\Sigma_{\tau}$ does not approach a curvature singularity as $\tau  \rightarrow $ 0. 

 We begin with the following general result.

\bbgin{theorem} \label{t 4.1.}
  Suppose $\sigma (\Sigma ) \leq $ 0 and (4.1) holds. Then each component $\Sigma_{o}$ of $\partial_{o}M_{{\cal F}}$ is a smooth, complete non-compact maximal hypersurface $\Sigma_{o} \subset {\bf M}$, weakly embedded in $\Sigma ,$ c.f. (1.27).
\end{theorem}

{\bf Proof:}
 Referring to the estimates in \S 2, by Proposition 2.2, $|K|^{2}$ is uniformly bounded on any compact subset $\Omega $ of {\bf M}
 and hence by (1.9),
\begin{equation} \label{e4.2}
inf_{K }\alpha  \geq  \alpha_{o} = \alpha_{o}(\Lambda_{o}, H_{o}) >  0, 
\end{equation}
i.e. there is no collapse of the lapse function on compact subsets. Further, we recall that $\alpha $ satisfies the Harnack inequality (2.21), i.e.
\begin{equation} \label{e4.3}
\frac{sup_{B(r_{o})}\alpha}{inf_{B(r_{o})}\alpha} \leq  C, 
\end{equation}
where $B(r_{o})$ is any geodesic $r_{o}$-ball in $(\Sigma_{\tau}, g_{\tau})$ and $C$ is independent of $\tau ,$ for $\tau\in [-\tau_{o},$ 0].

 The estimate (1.8) also gives an upper bound on $\alpha $ when $H$ is bounded away from 0, (as in (2.28)). However, when $H = \tau  \rightarrow $ 0, the lapse function $\alpha $ may well blow-up, i.e. diverge to $+\infty .$ Applying the Harnack inequality iteratively to a collection of balls covering a domain of bounded diameter, it follows that $\alpha $ either remains uniformly bounded on uniformly bounded domains within $\Sigma_{\tau},$ or $\alpha $ diverges uniformly to $+\infty $ on such domains, as $\tau  \rightarrow $ 0. 

  Given these preliminary remarks, let $\tau_{i}$ be any sequence with $\tau_{i} \rightarrow $ 0 monotonically and consider the behavior of $\{\Sigma_{\tau_{i}}\}.$ We divide the discussion into three cases.

 Suppose first that there is a constant $T_{o} <  \infty $ such that
\begin{equation} \label{e4.4}
sup_{\Sigma_{\tau_{i}}}t = t_{\tau_{i}} \leq  T_{o}, \ \ {\rm as} \ \ i \rightarrow  \infty . 
\end{equation}
As in the proof of Theorem 0.1, it then follows that $\Sigma_{\tau_{i}}$ converges smoothly to a compact maximal hypersurface $\Sigma_{o},$ and $\Sigma_{o}$ is diffeomorphic to $\Sigma .$ Of course, this situation implies $\sigma (\Sigma ) > $ 0, contradicting the assumption. (This case will be discussed further below).

 Next suppose
\begin{equation} \label{e4.5}
inf_{\Sigma_{\tau_{i}}}t = t_{min}(\tau_{i}) \rightarrow  \infty , \ \ {\rm as} \ \ i \rightarrow  \infty , 
\end{equation}
so that $t$ tends uniformly to $+\infty $ on $\Sigma_{\tau_{i}}$ as $i \rightarrow  \infty .$ Thus, the surfaces $\Sigma_{\tau_{i}}$ diverge uniformly to $\infty $ in infinite proper time. In this case, we then clearly have
$${\bf M} = M_{{\cal F}} $$
to the future of $\Sigma_{\tau_{o}},$ and so $\partial M_{{\cal F}} = \emptyset  $ again. 

 It follows that there are base points $y_{i}\in\Sigma_{\tau_{i}}$ such that $limsup_{i\rightarrow\infty} t(y_{i}) <  \infty ,$ and other points $z_{i}\in\Sigma_{\tau_{i}}$ where $liminf_{i\rightarrow\infty}t(z_{i}) = \infty .$ Analogous to the decomposition (3.21), for any $T_{o} <  \infty $ and any $i$, write 
\begin{equation} \label{e4.6}
\Sigma  = \Sigma_{\tau_{i}} = \Sigma^{T_{o}}\cup  \Sigma_{T_{o}}, 
\end{equation}
where $\Sigma^{T_{o}} = \{x\in\Sigma : t(x) \geq  T_{o}\}$, $\Sigma_{T_{o}} = \{x\in\Sigma : t(x) \leq T_{o}\}$. Hence for all $i$ and $T_{o}$ sufficiently large, both $\Sigma^{T_{o}}$ and $\Sigma_{T_{o}}$ are non-empty. Note that these domains need not be connected. Lemma 2.3 implies that the domains $\Sigma_{T_{o}} = (\Sigma_{i})_{T_{o}} \subset  (\Sigma_{\tau_{i}}, g_{\tau_{i}})$ do not collapse anywhere, i.e. at any sequence of base points. Further, as in the proof of Theorem 0.1, elliptic regularity applied to (2.2) gives uniform local control over the derivatives of $Ric$ and $K$ on $(\Sigma_{i})_{T_{o}}.$ It then follows exactly as in the proof of Theorem 0.1, that for any fixed $T_{o}$, and at any sequence of base points, the domains $(\Sigma_{i})_{T_{o}}$ have a subsequence converging smoothly and uniformly on compact subsets to a limit surface $(\Sigma_{o})_{T_{o}};$ the limit obtained of course depends on the choice of base points. As following (3.20), choose then a sequence $T_{j} \rightarrow  \infty $ and consider the double sequence $(i, j)$. A suitable diagonal subsequence of $(\Sigma_{i})_{T_{j}}$ then converges smoothly and uniformly on compact subsets to a limit $(\partial_{o}M_{{\cal F}}, g_{o}).$ The limit $\partial_{o}M_{{\cal F}}$ is a collection of maximal hypersurfaces in ({\bf M, g}) and forms the boundary of $M_{{\cal F}}$ (to the future of $\Sigma_{\tau_{o}}).$ Since each $\Sigma_{\tau_{i+1}}$ lies to the future of $\Sigma_{\tau_{i}},$ the limit $\partial_{o}M_{{\cal F}}$ is unique. By construction, the components $\Sigma_{o}$ of $\partial_{o}M_{{\cal F}}$ are weakly embedded in $\Sigma .$ Of course the complementary domains $\Sigma_{i}^{T_{j}}$ diverge to infinity and hence have no limit in ({\bf M, g}) itself.

{\endproof}

\bbgin{remark} \label{r 4.2.}
  {\rm Theorem 4.1 extends to the case $\sigma (\Sigma ) > $ 0 with only a minor modification, but for clarity we have separated the cases. Thus, suppose $\sigma (\Sigma ) > $ 0.

 The first case above, where (4.4) holds, gives a limit $\Sigma_{o}$ which is compact, and so diffeomorphic to $\Sigma .$ By Theorem 0.1, the foliation $M_{{\cal F}}$ then extends a definite amount to the future in CMC time, i.e. the range of $\tau $ extends a definite amount past 0 into ${\Bbb R}^{+}.$ Thus, as in Remark 2.7, either there exists a curvature singularity to the finite proper time future of $\Sigma_{o}$ or the CMC foliation $\Sigma_{\tau}$ extends to all values of $\tau\in [0, \infty ).$ For the second (4.5) and third cases (4.6), the proof of Theorem 4.1 proceeds in the same way. Of course the recollapse conjecture would imply that neither of these last cases actually occurs.}
\end{remark}

\bbgin{remark} \label{r 4.3.}
  {\rm We have not asserted that the lapse functions $\alpha_{i}$ on $\Sigma_{i}$ converge to the lapse function $\alpha $ on $\Sigma_{o}$ as $\tau_{i} \rightarrow $ 0 in Theorem 4.1 or Remark 4.2. If $\alpha_{i}(x_{i})$ remains uniformly bounded as $i \rightarrow  \infty ,$ then the lapse functions converge smoothly to the limit lapse function $\alpha $ on $\Sigma_{o},$ and $\alpha $ satisfies the lapse equation (1.6) on $\Sigma_{o}.$ We will call this the 'non-degenerate' situation.

 However, it is possible, (although probably only in very special situations), that $\alpha_{i} \rightarrow  \infty $ uniformly as $\tau_{i} \rightarrow $ 0 even under the bound (4.4). Thus suppose $\alpha_{i}(x_{i}) \rightarrow  \infty $ with $t(x_{i}) \leq  T_{o}, x_{i}\in\Sigma_{\tau_{i}}.$ Then exactly as in the proof of Theorem 3.4, c.f. (3.26), renormalize $\alpha_{i}$ by setting $\bar \alpha_{i} = \alpha_{i}/\alpha_{i}(x_{i}).$ As discussed there, using (4.3), the functions $\bar \alpha_{i}$ converge smoothly to a limit lapse $\bar \alpha_{\infty}$ satisfying, for $\alpha  = \bar \alpha_{\infty},$
\begin{equation} \label{e4.7}
-\Delta \alpha  + |K|^{2}\alpha  = 0. 
\end{equation}
This 'degenerate' situation occurs exactly when the limit maximal surface $\Sigma_{o}$ is only weak maximum for the volume functional and not a strict maximum. Thus, there is a deformation of $\Sigma_{o},$ namely the flow of the vector field $\bar \alpha_{\infty}\cdot  T,$ where $T$ is the unit normal, which preserves the volume of $\Sigma_{o}$ to first order.

 If $\Sigma_{o}$ is compact, (as in the case (4.4) when $\sigma (\Sigma ) > $ 0) then the maximum principle applied to (4.7) shows that $K =$ 0 and $\bar \alpha_{\infty} =$ const on $\Sigma_{o}.$ This situation does actually occur, for example in the Taub-NUT metric with $m =$ 0, c.f. [29, \S 5.8]. However, if $\Sigma_{o}$ is non-compact, (corresponding to the situation (4.6)), it is not clear that $K =$ 0.}
\end{remark}

 We make some further observations on the geometry of $\partial_{o}M_{{\cal F}}$ in ({\bf M, g}). By Remark 4.2, we may assume that each component $\Sigma_{o}$ is non-compact. First, the constraint equation (1.2) shows that $s \geq $ 0 everywhere on $\Sigma_{o},$ so that $g$ is a complete metric of non-negative scalar curvature on $\Sigma_{o}.$ Each component $\Sigma_{o}$ is a partial Cauchy surface for $M^{+},$ and $\partial_{o}M_{{\cal F}}$ is a Cauchy surface for $M^{+}.$

 The proper time $t(x)$ is a proper, unbounded exhaustion function on $\Sigma_{o},$ as is the lapse function $\alpha ,$ (or its renormalization $\bar \alpha_{\infty}$ in the degenerate case).

 Now suppose the curvature bound (2.1), i.e. 
\begin{equation} \label{e4.8}
|{\bf R}
| \leq  \Lambda_{o}, 
\end{equation}
holds globally on $M_{{\cal F}}$ to the future of $\Sigma_{\tau_{o}}$ and hence also on $\partial_{o}M_{{\cal F}}.$ (This assumption, which is automatic on compact subsets $\Omega $ of {\bf M}, has not been used in the discussion above in \S 4). The estimates (2.8) and (2.23) thus give a uniform bound on $|K|$ and $|\nabla log \alpha|$ on $M_{{\cal F}}.$ Hence, by integration of the $0^{\rm th}$ order Bel-Robinson estimate (2.34) from $\tau_{o}$ to 0, we obtain the bound
\begin{equation} \label{e4.9}
\int_{\partial_{o}M_{{\cal F}}}|Ric|^{2} + |dK|^{2} <  \infty . 
\end{equation}
The integrand $|Ric|^{2} + |dK|^{2}$ is of course pointwise bounded on $\partial_{o}M_{{\cal F}}$ by (2.8) and (1.7), (4.8). However, one certainly expects that each component $\Sigma_{o}$ is not volume collapsing anywhere, i.e.
\begin{equation} \label{e4.10}
vol B_{x}(1) \geq  v_{o} > 0, 
\end{equation}
for all geodesic balls $B_{x}(1) \subset  \Sigma_{o}$ and some $v_{o} > $ 0. If (4.10) does hold, then (4.9) implies that
\begin{equation} \label{e4.11}
|Ric| \rightarrow  0, |K| \rightarrow  0, 
\end{equation}
uniformly at infinity in the components $\Sigma_{o}.$ Hence the surfaces $\Sigma_{o}$ become flat and totally geodesic at infinity in this case.

\bbgin{remark} \label{r 4.4.}
  {\rm There are numerous interesting open questions concerning the structure of the components $\Sigma_{o}.$ For example, does $\Sigma_{o}$ admit a complete metric of uniformly positive scalar curvature, c.f. [26] for results on the structure of such manifolds. Do the components $\Sigma_{o}$ have at least two ends, and is there an essential 2-sphere $S^{2} \subset  \Sigma_{o}?$ Are the ends of $\Sigma_{o}$ asymptotically flat?

 It is also interesting to speculate if there is any relation of the structure of the manifolds $\Sigma_{o}$ weakly embedded in $\Sigma $ with the sphere or prime decomposition of $\Sigma .$ To explain this, as noted in \S 0, not every closed oriented 3-manifold admits a geometric structure in the sense of Thurston, nor does every 3-manifold admit a strong geometrization in the sense that (0.8) holds with incompressible ${\cal T} .$ The essential 2-spheres in $\Sigma $ are obstructions to the strong decomposition (0.8) and the geometrization conjecture [38] implies that these are the only obstructions, i.e. if $\Sigma $ has no essential 2-sphere, then $\Sigma $ admits a (unique) strong decomposition (0.8). Thus, one might speculate that a reducible 3-manifold $\Sigma ,$ i.e. a 3-manifold with an essential 2-sphere, cannot satisfy the assumptions of Theorem 0.3, so that in particular $\Sigma $ cannot evolve to the future for infinite proper time. Hence $\Sigma_{o}$ would be non-empty and it is then natural to consider the question if essential 2-spheres of $\Sigma $ are captured in the topology of $\Sigma_{o}.$}
\end{remark}

 We also make the following conjecture on the structure of the region to the future of $\partial_{o}M_{{\cal F}}$ in {\bf M}.

\medskip

{\bf Non-Compact CMC Conjecture.}
 If $\partial_{o}M_{{\cal F}} \subset $ {\bf M}
 is non-empty, then either there exists a CMC foliation ${\cal F}^{+}$ of $M^{+}$ by non-compact leaves $\Sigma_{\tau},$ defined for all $\tau\in (0,\infty ),$ and satisfying 
\begin{equation} \label{e4.12}
M_{{\cal F}^{+}} = \cup\Sigma_{\tau} = M^{+}, 
\end{equation}
or there is a curvature singularity to the finite proper-time future of $\partial_{o}M_{{\cal F}}$ in {\bf M}.

 This conjecture is a non-compact analogue of Theorem 0.1. Using the estimates in \S 2, together with the results and methods of [24], the conjecture can be proved provided the existence of suitable barrier surfaces $S_{i}$ can be established, i.e. surfaces in $M^{+}$ whose mean curvature $H(S_{i})$ diverges everywhere to $+\infty $ as $i \rightarrow  \infty .$ 

 Of course, one could consider a strengthening of this conjecture, asserting that there are no curvature singularities of {\bf M} reachable in finite CMC time, so that ${\cal F}^{+}$ covers all of $M^{+};$ this is the analogue of the CMC global existence problem in $M^{+}.$

 Note also that this conjecture implies the singularity formation conjecture of \S 0, since if there are no curvature singularities to the finite proper time future of $\partial M_{{\cal F}}$ then the leaves $\Sigma_{\tau}$ limit on a crushing singularity as $\tau  \rightarrow  \infty .$ This behavior must also occur at a finite proper future time from $\partial_{o}M_{{\cal F}},$ by the same argument given at the end of \S 2. Hence ({\bf M, g}) cannot be future geodesically complete.

\section{Remarks on the Curvature Assumption.}
\setcounter{equation}{0}

 In this section, we discuss the curvature assumption (0.11), which is the strongest assumption in Theorem 0.3. Of course one would like to replace this by the assumption that ({\bf M, g}) is merely future geodesically complete and derive (0.11) as a consequence. We will ignore the assumption on $|\nabla{\bf R}
|$ in (0.11), since it is of lesser importance.

 Let $q$ be any point in $\partial{\bf M},$ i.e. at the boundary of the maximal Cauchy development, (assumed not to be at infinity). Thus, near $q$, either the space-time ({\bf M, g}) is inextendible, for example because the curvature blows up, or ({\bf M, g}) is extendible as a space-time, (possibly not vacuum), near $q$ and so ({\bf M, g}) has a non-empty Cauchy horizon passing through $q$.

 For any point $x\in{\bf M},$ let $t_{\partial{\bf M}}(x)$ be the supremum of the times $t$ such that
\begin{equation} \label{e5.1}
B_{x}(t(x)) \subset  {\bf M}. 
\end{equation}
The condition (5.1) means that any past or future directed time-like geodesic starting at $x$, which has proper length less than $t_{\partial M}(x),$ has its endpoint in {\bf M}. Thus, $t_{\partial {\bf M}}(x)$ is essentially the proper time distance, to the past or future, to the time-like nearest point of $\partial {\bf M}$.

\medskip

{\bf Curvature Blow-up Problem.}
 Let ({\bf M, g}) be a CMC cosmological vacuum space-time. Does there exist a constant $C = C({\bf M, g}) <  \infty ,$ such that,
\begin{equation} \label{e5.2}
|{\bf R}| \leq  C/t_{\partial {\bf M}}^{2}. 
\end{equation}
As in (0.11), the estimate (5.2) is scale-invariant. Since the vacuum Einstein equations are also scale-invariant, the estimate (5.2) can be viewed as an upper bound on the rate of curvature blow-up on approach to a singularity, i.e. a point in $\partial{\bf M}
,$ or as an upper bound on the decay of the curvature in (large) distances from $\partial{\bf M}
.$ By scale invariance, these statements are equivalent. Of course the estimate (5.2) implies the curvature assumption (0.11) without the $|\nabla {\bf R}|$ bound, when ({\bf M, g}) is geodesically complete to the future of $\Sigma_{\tau_{o}}.$

 There are numerous examples where the opposite inequality holds, i.e. $|{\bf R}| \geq  c/t_{\partial {\bf M}}^{2},$ so that (5.2) cannot be sharpened significantly, except perhaps by removing the dependence of $C$ on the space-time ({\bf M, g}).

 We note that the estimate (5.2) has been proved for time-independent, i.e. stationary space-times, in [3], where $\partial{\bf M}$ is interpreted as the region where the associated Killing field becomes null.

 We first observe that (5.2) is in fact false for general globally hyperbolic space-times, i.e. without compact CMC Cauchy hypersurfaces. For suppose $({\bf M' , g'})$ is a geodesically complete globally hyperbolic vacuum space-time. For any such space-time, $\partial{\bf M'} = \emptyset  ,$ so that $t_{\partial {\bf M'}} = \infty$. Hence, (5.2) would imply that ${\bf R'} =$ 0, i.e. $({\bf M' , g'})$ is flat. 

  However, there exist such vacuum space-times which are not flat, namely the Christodoulou-Klainerman (CK) space-times arising as global perturbations of Minkowski space, [17]. Note that the CK space-times in fact have a global foliation by maximal hypersurfaces. There seem to be no other known examples and it is of basic importance to understand the general class of geodesically complete globally hyperbolic space-times.

 Now for cosmological space-times, $\partial{\bf M}$ must of course be non-empty by the Hawking-Penrose singularity theorem. However, the failure of the estimate (5.2) is closely related to the existence of the CK and any related space-times. Namely, if (5.2) were false, then there must exist a sequence of globally hyperbolic cosmological space-times $({\bf M}_{i}, {\bf g}_{i}),$ (possibly identical), and points $x_{i}\in{\bf M}_{i}$ with $t_{i}(x_{i}) \rightarrow $ 0, $t_{i}(\cdot  ) = t_{\partial{\bf M}_{i}}(\cdot ),$ (this may be arranged by a rescaling if necessary), such that the curvature at $x_{i}$ satisfies
\begin{equation} \label{e5.3}
|{\bf R}_{i}|(x_{i})\cdot  t_{i}(x_{i})^{2} \rightarrow  \infty . 
\end{equation}
We may assume without loss of generality that the points $x_{i}$ realize the maximal value of $|{\bf R}_{i}|$ on the level set $L_{i} = t_{i}^{-1}(x_{i}).$ By a standard argument, (the same as that establishing (3.39) but with time in place of space), we may choose the points $x_{i}$ to realize an approximate maximal value of $|{\bf R}_{i}|t_{i}^{2}$ in small time-like annuli about $L_{i}.$

 Next, rescale the metrics ${\bf g}_{i}$ by setting ${\bf g}_{i}'  = |{\bf R}_{i}|(x_{i})\cdot  {\bf g}_{i},$ so that 
\begin{equation} \label{e5.4}
|{\bf R}_{i}'|(x_{i}) = 1, 
\end{equation}
and consequently $|{\bf R}_{i}'| \leq  C <  \infty $ within ${\bf g}_{i}' $ bounded time distance to $x_{i}.$ This has the effect of making $t_{i}' ,$ the proper distance to the boundary w.r.t. $g_{i}' ,$ tend to $\infty .$ As discussed in \S 3, a subsequence then converges weakly in $L^{2,p}$ to a $L^{2,p}$ limit $({\bf M}
_{\infty}, {\bf g}
_{\infty}, x_{\infty}),$ (unwrapping in the case of collapse). The limit is necessarily a geodesically complete globally hyperbolic space-time. Further, if the condition (5.4), or some weakening of it, passes to the limit, then $({\bf M}
_{\infty}, {\bf g}
_{\infty})$ is not flat. (This is not at all obvious however, and may be difficult to prove).

 In addition, if $({\bf M}_{i}, {\bf g}_{i})$ are CMC cosmological space-times and $x_{i}\in (M_{{\cal F}})_{i},$ then the apriori bound (1.15) and the arguments from \S 3 show that the limit $({\bf M}_{\infty}, {\bf g}_{\infty})$ has global foliation by maximal hypersurfaces, as in the CK space-times.

 Thus, the problem is essentially whether such non-flat CK or related space-times can actually arise in this way, or more precisely whether the evolution from smooth Cauchy data can lead to singularities which have this blow-up character.

\medskip

 We conclude this paper with the following remark or caveat. The discussion above presents relations between the geometrization of 3-manifolds, (at least in case $\sigma (M) \leq $ 0), and the Einstein evolution equations. Thus, one may ask if a detailed study of the long-time behavior of the hyperbolic Einstein equations is a possible avenue of approach to the solution of the geometrization conjecture. And conversely, if the resolution of the geometrization conjecture has any direct bearing on the resolution of some of the fundamental problems on global existence and asymptotics for the Einstein evolution. 

 With regard to the first problem, this seems to be an extremely difficult approach. Approaches to the solution of the geometrization conjecture by study of a suitable sequence of {\it elliptic} PDEs are given in [1], [2] for example, or by study of a {\it parabolic} PDE, the Ricci flow, in [27] for example. Both the elliptic and the parabolic approaches appear to be much simpler than the hyperbolic approach via the Einstein flow and stand a much better chance of resolving geometrization. For instance, the analogue of Theorem 0.3 has been proved in both of these approaches, c.f. [1], [2] and [28], with only the bound on {\bf R} in (0.11), i.e. without any bound on $|\nabla {\bf R}|$. Further, strong geometrization in the sense of Definition 0.2 has been proved in this case in these approaches.

 Similarly, it is not at all clear that the (affirmative) resolution of the geometrization conjecture would have any concrete implications on the resolution of any of the global existence, asymptotics, or singularity formation problems in general relativity. Nevertheless, it is clear that there are many interesting relations between these two subjects.

\bibliographystyle{plain}

\bigskip
\begin{center}
June 2000
\end{center}
\medskip
\address{Department of Mathematics\\
S.U.N.Y. at Stony Brook\\
Stony Brook, N.Y. 11794-3651\\}
\email{anderson@@math.sunysb.edu}

\end{document}